# Electrochemical Impedance of a Battery Electrode with Anisotropic Active Particles

Juhyun Song[a,1,2] and Martin Z. Bazant[a,b,1,*]

[a]Department of Chemical Engineering, Massachusetts Institute of Technology,

77 Massachusetts Avenue, Cambridge, Massachusetts 02139, United States

[b]Department of Mathematics, Massachusetts Institute of Technology,

77 Massachusetts Avenue, Cambridge, Massachusetts 02139, United States

---

[1] ISE member

[2] Current address

    Samsung Cheil Industries Inc., Research Center, 332-2 Gocheon-dong, Uiwang,

    Gyunggi 437-711, South Korea

[*] Corresponding author

    Postal address: 25 Ames Street, Room 66-552, Cambridge, MA 02139, United States

    E-mail address: bazant@mit.edu

    Tel.: +1 617 258 7039

**Abstract**

Electrochemical impedance spectra for battery electrodes are usually interpreted using models that assume isotropic active particles, having uniform current density and symmetric diffusivities. While this can be reasonable for amorphous or polycrystalline materials with randomly oriented grains, modern electrode materials increasingly consist of highly anisotropic, single-crystalline, nanoparticles, with different impedance characteristics. In this paper, analytical expressions are derived for the impedance of anisotropic particles with tensorial diffusivities and orientation-dependent surface reaction rates and capacitances. The resulting impedance spectrum contains clear signatures of the anisotropic material properties and aspect ratio, as well as statistical variations in any of these parameters.

# 1. Introduction

Electrochemical impedance spectroscopy (EIS) is used in various fields, such as energy storage and conversion[1-5], cell biology[6, 7], corrosion science[8, 9], and catalysis[10, 11], to characterize transport, reaction, and accumulation of charge carriers in the systems. For insertion battery electrodes, it has also been widely used across many different material compositions[1, 2, 12] and presented us useful information for understanding the mechanistic journey of charge carriers, including ion transport through electrolyte solution and a porous separator, accumulation and insertion reaction on electrolyte/active material interface, and solid-state diffusion in active particles. In a typical impedance spectrum of a battery electrode[13, 14], the response approaches a purely resistive behavior at the highest frequency limit (> 10 kHz), which corresponds to the transport resistance of ions through bulk electrolyte solution and a separator. Subsequently in the high frequency range (10 kHz ~ 10 Hz), the impedance response gradually transitions from a capacitive behavior to a resistive behavior as the applied frequency decreases, which has been widely described by various resistive-capacitive (RC) circuit models. It represents the behavior of charge carriers at the electrolyte/active material interface; most of them are being accumulated around the interface when the applied frequency is higher than the transition frequency, and are participating in the ion insertion reaction when the applied frequency is lower. At the low frequencies (< 10 Hz), the impedance has a bounded diffusion (BD) element which looks like a hockey stick in its complex plane plot. It corresponds to the solid-state transport of charge carriers in active particles. In the BD element, a Warburg regime is observed with near-45° slope when the applied frequency is higher than the diffusion characteristic frequency and the ion penetration depth from the interface is shorter than the particle length scale. On the other hand, as the penetrations reach the centers of particles at the lower frequencies, the BD element transitions to a capacitive regime, drawing an almost vertical line with diverging magnitude[15-18]. An impedance spectrum in reality may look much more complicated as it may include contributions from a solid electrolyte interface (SEI) layer[19-21], a phase transformation of active material[22], and an additional conductive/insulating coating on active particles[23].

To interpret such impedance spectra and extract meaningful parameters, we need a model that appropriately accounts for essential characteristics of the electrode. Many impedance models have been introduced to describe behavior of charge carriers in each component of a battery electrode. Their behavior on the electrolyte/active material interface and in the active particles is usually described using the Randles model[15, 24, 25]. It involves an ideal capacitor or a constant-phase element (CPE), which describes accumulation of charge carriers on the interface, and a series of a resistance and a Warburg-type element, which represents the ion insertion reaction followed by the solid-state transport in active particles[26-28]. Electrolyte solution for battery cells always contains an excess amount of supporting electrolyte in order to achieve high conductivity. Therefore, the transport of ions in electrolyte solution and a porous separator can simply be modeled by a series resistance in the typical frequency range of impedance measurement[13, 14]. For porous composite electrodes, there are three simultaneous processes occurring in the same macroscopic phase: ion transport through electrolyte phase, electron conduction through conductive phase, and the electrochemical processes involving active particles. Their impedance models have also been developed based on macroscopically homogenized conduction and reaction models[20, 21, 29, 30], and they can be represented by a distributed transmission line that hierarchically involves a sub-circuit for the electrochemical processes taking place at the interface and in the active particles. The contribution from ion and electron conduction along the electrode thickness, however, would be negligible when the conductivities are high and/or the electrode is thin enough[17]. Although many impedance models have been suggested in the community to account for various aspects of battery electrodes, most of them still assume isotropic properties for the active particles, regardless of their scope of interest.

However, most battery materials currently under investigation are strongly anisotropic, which means that behavior of charge carriers in the materials varies with the direction in respect to the crystallographic axes[31-37]. In principal, the anisotropy is attributed to different activation barriers along the hopping paths of charge carriers depending on the direction. Many battery materials have high electron mobility compared to that of ions[18, 32]. Others could be modified to increase electron conductivity through cation doping[38-41] and conductive coating[42]. Electrons can therefore be considered freely available throughout the active particles, and it is the ion hopping that determines the anisotropy in transport of charge carriers as well as in surface insertion kinetics [18, 43, 44]. For example, $Li_xCoO_2$ has a layered metal oxide structure in Figure 1 (a)[45], where lithium ions can move quickly through the plane between the metal oxide layers, but their movement across the layers is less likely and very slow[33, 34]. On the other hand, $Li_xFePO_4$ has an olivine structure in Figure 1 (b)[45], where lithium ions can move quickly through one-dimensional channels in the $b$-crystallographic direction[35-37]. Models also predict that intercalation kinetics[46], phase separation dynamics[47], and nucleation[48] are highly anisotropic due to tensorial coherency strain and different composition-dependent surface on each crystal facet[49, 50]. Like $Li_xFePO_4$, other important battery materials tend to phase-separate when they are alloying with Li ions[51-53], and their impedance characteristics are beginning to be considered[22]. In this paper, however, we confine our scope to the materials forming a single phase solid solution to focus on the effects of anisotropic properties. This also includes materials that tend to phase-separate, such as $Li_xMn_2O_4$ and $Li_xFePO_4$, while outside of their miscibility gaps for $x \approx 0$ or $x \approx 1$.

Isotropic models of active particles are widely employed in EIS studies of batteries, in part because traditional active particles were large enough to have many randomly oriented crystal grains and presented lumped isotropic behaviors[54], but in such cases, only apparent electrochemical parameters could be obtained rather than intrinsic properties of the active material. Meanwhile, with the recent developments in synthesis of active materials, modern battery electrodes are made of small particles with a few grains or even a single grain, to achieve excellent rate capability and long cycle life[55-58]. Then, the anisotropic particle properties may significantly affect their impedance spectra, and this motivates us to consider the effects of particle anisotropies on battery impedance. Therefore in this study, we first illustrate a generic approach to arrive at an impedance model for a battery electrode that has anisotropic active particles. Using the generic framework, we then analytically study the impedance of an anisotropic 2D rectangular particle, where we could investigate the effects of anisotropies in diffusion, surface kinetics, and surface capacitance. Lastly, we study the overall impedance of a battery electrode when the particles have anisotropic distributions in length scales. Throughout this work, we assumed that the active material forms a homogeneous solid solution which has high electron mobility. It is also assumed for a porous electrode that its thickness is nominal and the electrolyte conductivity is high so that the models do not account for any gradient that may develop along the electrode thickness. When such a gradient becomes important, the models presented here can be incorporated into the impedance models of thick porous electrodes proposed in References 20 and 30.

## 2. General Theory

EIS is complex generalization of Ohmic resistance in the frequency domain, and it includes information about phase difference as well as relative magnitudes of perturbations in potential and current[14]. It is measured by applying a small perturbation, either in potential of current, to an electrochemical system around a reference state and measuring the response in the other variable. Impedance can be properly defined for a linear time invariant (LTI) system by a transfer function of the potential perturbation over the current perturbation, given that the causality between the stimulus and the response is obeyed[59]. A linear integral transformation, usually the Fourier or the Laplace transformation, is employed in defining the transfer function. For most real electrochemical systems including a battery electrode, however, there are inherent nonlinearities due to complex reaction kinetics and non-ideal transports in condensed phases. It can therefore be properly defined only when the amplitudes of the perturbations are small enough to approximate the system mathematically linear.

While any arbitrary form of stimulus could be applied, a monochromatic sinusoidal perturbation is usually employed in EIS practices, sweeping the applied frequency within a certain range. When a system is approximated linear and perturbed by a small sinusoidal stimulus, relevant variables fluctuate correspondingly in isofrequential sinusoidal forms with negligible harmonic responses. When applying a regular perturbation analysis, it is therefore enough to consider the system response up to the first order in perturbation. In the analysis, an arbitrary system variable, $X$, can be expanded as a power series of a sufficiently small parameter, $\varepsilon$, where the zero-order term and the $\varepsilon$-order term represent the reference state response and the sinusoidal perturbation, respectively. The higher-order terms, represented by the mathematical order of $\varepsilon^2$, correspond to the non-linear harmonic responses, which are relatively small.

$$X = X_0 + \varepsilon X_1 + O(\varepsilon^2) = X_0 + \varepsilon \hat{X} \exp(i\omega t) + O(\varepsilon^2) \tag{1}$$

where $i = \sqrt{-1}$ is the unit imaginary number, $\omega$ is the radial frequency, and $t$ is the time variable. The Fourier transform of the perturbation term, $X_1$, yields $\hat{X}$, the Fourier coefficient, which is a complex number containing information related to the magnitude and the phase of the perturbation in $X$.

*2.1. Anisotropic transport of charge carriers*

The system under initial investigation is a single crystal nanoparticle of anisotropic battery material, in which diffusivities of charge carriers vary depending on the diffusion direction. The system is fully exposed to electrolyte solution, except small contacts to electron conducting material. Ions intercalate into the system from its interface with electrolyte solution, and electrons come from its contacts with conducting material. In many battery materials, the mobility of electrons is much higher than that of ions[18, 32], and electrons may be considered freely available in the system. Ions are quickly surrounded and shielded by electrons due to the electrostatic driving force, and the local charge neutrality is achieved in the macroscopic point of view. Under such conditions, the neutral diffusion of ions limits the solid-state transport of charge carriers, and the Fick's neutral diffusion equation can be recovered in its tensor form, to describe the anisotropic ion transport in the system.

$$\frac{\partial c}{\partial t} = \nabla \cdot (\mathbf{D}_{ch} \nabla c) \tag{2}$$

where $c$ is the local ion concentration, and $\mathbf{D}_{ch}$ is the chemical diffusivity tensor. It is always possible to make $\mathbf{D}_{ch}$ diagonal by setting the coordinate axes aligned with the diffusion principal axes, along which diffusion is not influenced by concentration gradient in the other directions.

$$\mathbf{D}_{ch} = \begin{pmatrix} D_{ch,x} & 0 & 0 \\ 0 & D_{ch,y} & 0 \\ 0 & 0 & D_{ch,z} \end{pmatrix} \quad (3)$$

Each component of $\mathbf{D}_{ch}$ is the diffusivity in corresponding direction. For orthorhombic crystal structures at least, including spinel, olivine, and some of layered metal oxides, the diffusion principal axes coincide with the crystallographic axes. Still, the diffusion is generally not isotropic and the diagonal components may be different from the others. For olivine structure as an example, $D_{ch,y}$ is much larger than the two others, and Li ions have effectively 1D transport in $y$ or (0 1 0) direction[32, 35]. Calculation of the ion diffusivities in battery materials has been extensively studied in the community using the first principle approaches[32, 34, 35, 60], and it has been shown that the anisotropy in diffusion depends on the system size[36].

In calculation of impedance, the anisotropic diffusion equation should be expanded by substituting the power series expression in Equation (1):

$$\frac{\partial c_0}{\partial t} + \varepsilon \frac{\partial c_1}{\partial t} + O(\varepsilon^2) = \nabla \cdot \left( \left( \mathbf{D}_{ch,0} + \varepsilon \mathbf{D}_{ch,1} + O(\varepsilon^2) \right) \nabla \left( c_0 + \varepsilon c_1 + O(\varepsilon^2) \right) \right) \quad (4)$$

Notice $\mathbf{D}_{ch}$ may also fluctuate, since it is a function of $c$ due to any non-ideal interaction between the ions and the host material. The expanded terms can be collected according to their orders in $\varepsilon$, to give a set of linear differential equations to be successively solved. When the system is perturbed around an uniform reference state, the zero-order problem is trivial and results in the gradient-free solution, where $\nabla c_0 = \mathbf{0}$. The $\varepsilon$-order problem, by which the impedance response is defined, can then be reduced to an ordinary linear differential equation in the frequency-space domain through Fourier transformation.

$$i\omega \hat{c} = \nabla \cdot (\mathbf{D}_{ch} \nabla \hat{c}) \qquad (5)$$

Hereby we denote the chemical diffusivity tensor evaluated at the reference state simply by $\mathbf{D}_{ch}$, instead of $\mathbf{D}_{ch,0}$. Likewise, for any Fourier transformed equations in this article, parameters without the Fourier's hat notation ($\hat{\square}$) represent their values evaluated at the reference state. Note that the $\mathbf{D}_{ch,1}$ term would survive in the $\varepsilon$-order problem, if the reference state has non-uniform composition. Given the linearity approximation is valid, the higher-order solutions are negligible compared to the $\varepsilon$-order solution, and Equation (5) governs the transport of charge carriers in the system.

Transport of charge carriers in active material has been described by current flow in a distributed RC transmission line[61, 62]. Isotropic transport in traditional approach could be modeled using a 1D transmission line shown in Figure 2 (a). Likewise, anisotropic transport can be modeled using a 2D or a 3D transmission line shown in Figure 2 (b) and (c), where the resistances between the nodes have different values depending on the direction. In the equivalent circuits, grids in black represent ion pathways, and grids in grey represent electron pathways. When the high electron mobility was assumed, the electron pathways (in grey) have negligible resistance between nodes. Therefore, there is no electric potential drop between the nodes, and the grids in grey become equipotential. On the other hand, potential and current in the black grids can be defined by the electrical equivalent of the chemical potential and the flux of ions, respectively. The resistance represents drag that ions experience when they transport, and the resistance values are direction-dependent when ion transport is anisotropic. For the edges of the circuits, various terminating elements may be employed depending on the boundary conditions. Such circuit representation allows relatively simple treatments based on the standard circuit theory, and we can modify and reformulate models according to the characteristics of the system.

*2.2. Anisotropic surface kinetics*

To focus on the effects of anisotropy in surface kinetics, we consider a simple interface model, where the active material is in direct contact with the electrolyte solution without any additional resistive layer, such as SEI layer. One of the electrochemical processes taking place on the electrolyte/active material interface is insertion of ions from the electrolyte solution into the active material. The insertion kinetics has been widely modeled by the Butler-Volmer equation, assuming even activeness over the entire particle surface. However, as discussed in Introduction, the insertion rate would be different according to the surface orientation with respect to the crystallographic axes. To be able to account for the anisotropy in surface kinetics, we make the charge transfer resistance in the model a function of surface orientation.

$$j_{ins} = j_0(\mathbf{n})\left[\exp\left(-\alpha(\mathbf{n})\frac{e\eta}{kT}\right) - \exp\left((1-\alpha(\mathbf{n}))\frac{e\eta}{kT}\right)\right] \qquad (6)$$

where $j_{ins}$ and $j_0(\mathbf{n})$ are the insertion current density and the exchange current density, respectively. $e$ is the electron charge constant, $\mathbf{n}$ is the surface normal vector, $\alpha(\mathbf{n})$ is the transfer coefficient $(0 < \alpha < 1)$, and $\eta = \Delta\phi - \Delta\phi_{eq}$ is the surface overpotential. Notice that $j_0$ and $\alpha$ are now functions of the surface normal vector, $\mathbf{n}$.

Applying the regular perturbation analysis, variables in the Butler-Volmer equation can be expanded using the power series expression in Equation (1). We can collect $\varepsilon$-order terms to calculate the impedance response for the insertion reaction. When the system is perturbed around an equilibrium reference state, fluctuations in $j_0$ and $\alpha$ do not effectively contribute to the impedance response, since the two exponential terms evaluated at the reference state balance out each other. On the other hand, perturbation in $\Delta\phi_{eq}$ brings the contribution of surface ion concentration. Taking the Fourier transformation, perturbation in $j_{ins}$ can be expressed by the following equation.

$$\hat{j}_{ins} = \frac{1}{\rho_{ct}(\mathbf{n})}\left(\Delta\hat{\phi} - \left(\frac{\partial\Delta\phi_{eq}}{\partial c}\right)\hat{c}_s\right) \qquad (7)$$

where $\rho_{ct}(\mathbf{n}) = kT/(j_0(\mathbf{n})e)$ is the charge transfer resistance and $\partial\Delta\phi_{eq}/\partial c$ is the Nernst shift, both evaluated at the reference state. $\hat{c}_s$ is the surface ion concentration, which may have different values depending on position on the surface. In more general, $\Delta\phi_{eq}$ as well as $\partial\Delta\phi_{eq}/\partial c$ may also depend on surface orientation, but we leave these subtle effects out of consideration and focus on $\rho_{ct}(\mathbf{n})$ only as a measure of anisotropy in surface kinetics.

The Faraday's law can be applied at the electrolyte/active material interface to correlates $j_{ins}$ and the ion flux in the active material. The insertion current can then be represented as:

$$\hat{j}_{ins} = -e\left(\mathbf{D}_{ch}(\nabla\hat{c})_s\right)\cdot\mathbf{n} \qquad (8)$$

where $(\nabla\hat{c})_s$ is the ion concentration gradient at the surface. When it is used together with the Butler-Volmer model in Equation (7), they present a boundary condition that governs the ion transport in active material.

$$-e\left(\mathbf{D}_{ch}(\nabla\hat{c})_s\right)\cdot\mathbf{n} = \frac{1}{\rho_{ct}(\mathbf{n})}\left(\Delta\hat{\phi} - \left(\frac{\partial\Delta\phi_{eq}}{\partial c}\right)\hat{c}_s\right) \qquad (9)$$

With this boundary condition, the anisotropic diffusion equation, Equation (5), can be integrated twice to give the ion concentration field in the system. Here, we neglected surface diffusion of charge carriers, which can be important in other studies[63].

*2.3. Anisotropic interface capacitance*

Another electrochemical process that takes place on the interface is accumulation of charge carriers. They would form an electrical double layer or directly adsorb onto the surface. One of the simplest models that describe such behaviors is the ideal capacitor model, where the amount of charge accumulation on the interface is proportional to the potential drop across it. The proportional constant is capacitance. The surface capacitance may have different values depending on the surface orientation with respect to the crystallographic axes. Capacitance from diffuse double layer should be fairly isotropic, although compact layer contributions (*e.g.* surface-specific ion adsorptions) will depend on surface orientation. On the other hand, pseudocapacitance, if there is any, should depend on surface orientation as it arises from surface-specific side reactions. To be able to account for the anisotropy in surface accumulation, we make the surface capacitance in the ideal capacitor equation a function of the surface orientation.

$$j_{acc} = C_{surf}(\mathbf{n}) \frac{d\Delta\phi}{dt} \tag{10}$$

where $j_{acc}$ is the accumulation current density on the interface, and $C_{surf}(\mathbf{n})$ is the surface capacitance that depends on the surface normal vector. Applying the regular perturbation analysis and the Fourier transformation, the ideal capacitor equation can be forged for impedance calculation in a straightforward manner:

$$\hat{j}_{acc} = i\omega C_{surf}(\mathbf{n}) \Delta\hat{\phi} \tag{11}$$

Other models of the surface accumulation, including the CPE model, could also be generalized to account for the anisotropy in surface capacitance, by making the relevant parameters depend on the surface orientation.

*2.4. Definitions of impedance functions*

Once we have anisotropic models for bulk transport, surface reaction and accumulation of charge carriers, we are now able to define impedance functions for each of the electrochemical processes. To begin with diffusion impedance, the contribution from anisotropic diffusion of ions in active material appears through the equilibrium potential of the insertion reaction, $\Delta\phi_{eq}$, since it is a function of the ion concentration, $c$. Therefore, the local diffusion impedance is defined with the partial derivative of $\Delta\phi_{eq}$ with respect to $c$:

$$z_D = \frac{\Delta\hat{\phi}_{eq}}{\hat{j}_{ins}} = \left(\frac{\partial \Delta\phi_{eq}}{\partial c}\right)\frac{\hat{c}_s}{\hat{j}_{ins}} \tag{12}$$

where $z_D$ is the local diffusion impedance. Employing this definition, Equation (7) can be rearranged into a generalized Ohm's form, which leads to the definition of local insertion impedance.

$$z_{ins} = \frac{\Delta\hat{\phi}}{\hat{j}_{ins}} = \rho_{ct}(\mathbf{n}) + z_D \tag{13}$$

where $z_{ins}$ is the local insertion impedance. It implies a circuit analogy of the ion insertion process, a series circuit of $\rho_{ct}$ and $z_D$, in a small perturbation regime. On the other hand, the impedance function corresponding to the charge carrier accumulation on the surface can be defined independently, from Equation (11).

$$z_{acc} = \frac{\Delta \hat{\phi}}{\hat{j}_{acc}} = \frac{1}{i\omega C_{surf}(\mathbf{n})} \tag{14}$$

where $z_{acc}$ is the local accumulation impedance. Since we employed the ideal capacitor model for surface accumulation, its impedance has the same form with an ideal capacitor. Depending on the employed model, $z_{acc}$ can be represented by a CPE or one of other models, too.

Given the impedance functions defined for each of the electrochemical processes, we can now move onto the combined response that these processes provide together in a battery electrode. We can define three different domain scales in studying the combined response, expanding our view from an infinitesimal local area, through a particle, to the entire electrode, as shown in Figure 3. The total local impedance is defined by the combined response of the electrochemical processes on the infinitesimal area on active particle surface. It tells us the activeness of a specific position on the particle surface. To define it, we assume independent parallel contributions of $j_{ins}$ and $j_{acc}$ at the particle surface, following Randles and Graham[24, 25].

$$\hat{j}_{tot} = \hat{j}_{ins} + \hat{j}_{acc} \tag{15}$$

where $j_{tot}$ is the total current density on the active particle surface. The total local impedance then can be defined by a transfer function of the potential over the total current density.

$$z_{tot} = \frac{\Delta \hat{\phi}}{\hat{j}_{tot}} = \frac{\Delta \hat{\phi}}{\hat{j}_{ins} + \hat{j}_{acc}} = \left(z_{ins}^{-1} + z_{acc}^{-1}\right)^{-1} = \left(\left(\rho_{ct}(\mathbf{n}) + z_D\right)^{-1} + z_{acc}^{-1}\right)^{-1} \tag{16}$$

where $z_{tot}$ is the total local impedance. As implied by the expression, $z_{tot}$ can be represented by the Randles equivalent circuit, which has $C_{surf}$ in parallel with a series of $\rho_{ct}$ and $z_D$.

Then, the particle impedance is defined by the combined response of the electrochemical processes on the surface of a single particle. The definition therefore employs the total current on the particle surface, which can be obtained by integrating $j_{tot}$ over the particle surface area. The potential is assumed homogeneous on the entire particle surface due to high electron mobility. Then, the particle impedance can be defined as:

$$Z_p = \frac{\Delta \hat{\phi}}{\hat{J}_p} = \frac{\Delta \hat{\phi}}{\int_{A_p} \hat{j}_{tot} dS} = \left( \int_{A_p} z_{tot}^{-1} dS \right)^{-1} \tag{17}$$

where $Z_p$ is the particle impedance, $J_p$ is the total current on the particle surface, and $A_p$ is the particle surface area. $Z_p$ can be considered as a harmonic average of the local impedance over the particle surface. This does make sense in the electric circuit point of view, because the particle response comes from the local responses of infinitesimal areas which are arranged in parallel or side by side to constitute the entire particle surface. By looking at the impedance response of a single particle, $Z_p$, we can investigate the effect of anisotropic parameters on impedance behavior of a battery electrode.

Expanding our view once again to the overall electrode, we can also define the overall electrode impedance using the combined response of the electrochemical processes on the overall surface of active particles in an electrode. The definition employs the overall current summed over the particle population. When some particle properties have variations and distributed in a variance, it is useful to convert the summation to an integral over the distributions. It can be done by defining continuous random variables that represent the distributed particle properties.

$$Z_{ov} = \frac{\Delta \hat{\phi}}{\hat{J}_{ov}} = \frac{\Delta \hat{\phi}}{\sum_{n_p=1}^{N_p} \hat{J}_{p,n_p}}$$
$$= \frac{\Delta \hat{\phi}}{N_p \int_{\Omega} \Pr_{\mathbf{V}_d}(\mathbf{v}_d) \hat{J}_p(\mathbf{v}_d) d\mathbf{v}_d} = N_p^{-1} \left( \int_{\Omega} \Pr_{\mathbf{V}_d}(\mathbf{v}_d) Z_p^{-1}(\mathbf{v}_d) d\mathbf{v}_d \right)^{-1}$$
(18)

where $Z_{ov}$ is the overall electrode impedance, $\hat{J}_{ov}$ is the overall electrode current, $n_p$ and $N_p$ are the numbering index and the number of active particles in the electrode, respectively. When $Z_{ov}$ is defined with the integral over distributed particle parameters, $\mathbf{v}_d$ is the realization of the random variable vector, $\mathbf{V}_d$, where each component represents one of the distributed parameters. $\Pr_{\mathbf{V}_d}$ is then the joint probability density function (PDF) of $\mathbf{V}_d$, and $\Omega$ is the domain in which $\mathbf{V}_d$ is defined. $\mathbf{V}_d$ may have diffusivities, resistances, and lengths in different directions for its components. Arriving at Equation (18), we assumed that the ion conductivity in the electrolyte solution is high and/or the electrode is thin enough, so that any gradient along the thickness direction was not considered here. It the assumptions are found invalid and there is a significant gradient along the thickness direction, the impedance definitions presented here can be incorporated into the impedance models of thick porous electrodes suggested in References 20 and 30. This definition of $Z_{ov}$ allows us to study the distribution effects of anisotropic particle properties. After all, to calculate the impedance functions derived above, we essentially need information about the ion concentration and their movements in the particles.

## 3. Impedance of an Anisotropic 2D Rectangular Particle

In this section, we focus on the impedance behavior of an anisotropic 2D rectangular particle shown in Figure 4. It can be considered as a hypothetic 2D particle or a rectangular cross-section of a rod-shaped particle, which is fully submerged in the electrolyte solution with small contacts to conductive materials. This system was chosen because we can fully study the effects of anisotropies on the impedance behavior with minimal complexity. However, the general models presented in the previous section can be applied to a battery electrode with any configuration. To obtain impedance behavior of the system, we need to solve the ion transport problem. The ion transport problem in the system is reduced to a 2D boundary value partial differential equation (PDE) problem, for which we can obtain an analytical solution. The governing diffusion equation becomes

$$i\omega \hat{c} = D_{ch,x} \frac{\partial^2 \hat{c}}{\partial x^2} + D_{ch,y} \frac{\partial^2 \hat{c}}{\partial y^2} \tag{19}$$

The boundary conditions from the Butler-Volmer model and the system symmetry become:

$$-eD_{ch,x} \frac{\partial \hat{c}}{\partial x}\bigg|_{x=l_x} = \frac{1}{\rho_{ct,x}} \left( \Delta\hat{\phi} - \left( \frac{\partial \Delta\phi_{eq}}{\partial c} \right) \hat{c}\big|_{x=l_x} \right)$$

$$-eD_{ch,y} \frac{\partial \hat{c}}{\partial y}\bigg|_{y=l_y} = \frac{1}{\rho_{ct,y}} \left( \Delta\hat{\phi} - \left( \frac{\partial \Delta\phi_{eq}}{\partial c} \right) \hat{c}\big|_{y=l_y} \right)$$

$$\frac{\partial \hat{c}}{\partial x}\bigg|_{x=0} = 0$$

$$\frac{\partial \hat{c}}{\partial y}\bigg|_{y=0} = 0 \tag{20}$$

where $\rho_{ct,x} = \rho_{ct}(\mathbf{e}_x)$ and $\rho_{ct,y} = \rho_{ct}(\mathbf{e}_y)$ are the charge transfer resistances on $x$ and $y$ normal surfaces, respectively. $l_x$ and $l_y$ are the one halves of the lengths in corresponding directions, as shown in Figure 4.

The variables in Equations (19) and (20) can be nondimensionalized using appropriate scales. The dimensionless concentration can be defined by $\hat{\tilde{c}} = \left(-\partial \Delta \phi_{eq}/\partial c\right)\hat{c}/\Delta\hat{\phi}$, and the dimensionless spatial variables by $\tilde{x} = x/l_x$ and $\tilde{y} = y/l_y$. Further scaling becomes simpler when we define frequency scales and resistance scales for diffusion in each direction:

$\omega_{D,x} = D_{ch,x}/l_x^2$, $\omega_{D,y} = D_{ch,y}/l_y^2$, $\rho_{D,x} = \left(-\partial \Delta \phi_{eq}/\partial c\right)l_x/eD_{ch,x}$, and

$\rho_{D,y} = \left(-\partial \Delta \phi_{eq}/\partial c\right)l_y/eD_{ch,y}$, where $\omega_{D,x}$ and $\omega_{D,y}$ are the diffusion characteristic frequencies in $x$ and $y$ directions, and $\rho_{D,x}$ and $\rho_{D,y}$ are the diffusion characteristic resistances in $x$ and $y$ directions, respectively. Then, it turns out that four dimensionless numbers govern the ion transport in the system. Two are the ratios of frequency scales: $\tilde{\omega} = \omega/\omega_{D,x}$, $\tau = \omega_{D,y}/\omega_{D,x}$. The other two are the ratios of diffusion characteristic resistance and charge transfer resistance on each surface: $\beta_x = \rho_{D,x}/\rho_{ct,x}$ and $\beta_y = \rho_{D,y}/\rho_{ct,y}$. Each resistance ratio indicates relative insertion rate on the surfaces and diffusion rate in the bulk system, like a Biot number in traditional transport systems involving boundary fluxes. Employing the dimensionless variables and the scaled parameters, the governing equation and boundary conditions become

$$i\tilde{\omega}\hat{\tilde{c}} = \frac{\partial^2 \hat{\tilde{c}}}{\partial \tilde{x}^2} + \tau \frac{\partial^2 \hat{\tilde{c}}}{\partial \tilde{y}^2}$$

$$-\beta_x^{-1} \left.\frac{\partial \hat{\tilde{c}}}{\partial \tilde{x}}\right|_{\tilde{x}=1} = 1 + \left.\hat{\tilde{c}}\right|_{\tilde{x}=1}$$

$$-\beta_y^{-1} \left.\frac{\partial \hat{\tilde{c}}}{\partial \tilde{y}}\right|_{\tilde{y}=1} = 1 + \left.\hat{\tilde{c}}\right|_{\tilde{y}=1} \quad (21)$$

$$\left.\frac{\partial \hat{\tilde{c}}}{\partial \tilde{x}}\right|_{\tilde{x}=0} = 0$$

$$\left.\frac{\partial \hat{\tilde{c}}}{\partial \tilde{y}}\right|_{\tilde{y}=0} = 0$$

Such a boundary value PDE problem, defined in a finite domain, can be solved analytically by the finite Fourier transformation (FFT) method[64]. Mathematical works involved in deriving the solution, including the FFT method, are presented in Appendix B. The solution for $\hat{\tilde{c}}$ is given by

$$\hat{\tilde{c}}(\tilde{x},\tilde{y}) = -1 + \sum_{k=1}^{\infty} \Gamma_k \left(1 - \frac{\tilde{\rho}_{Dy/cty} \cosh(\Lambda_k \tilde{y})}{\tilde{\rho}_{Dy/cty} \cosh(\Lambda_k) + \Lambda_k \sinh(\Lambda_k)}\right) B_k \cos(\lambda_k \tilde{x}) \quad (22)$$

where $\lambda_k$, $B_k$, $\Lambda_k$, and $\Gamma_k$ are defined in Appendix B for $k = 1, 2, 3, \ldots$.

The particle impedance includes parallel contributions from charge accumulation on the surface and the ion insertion into the particle. Under such an arrangement, a resistive-capacitive (RC) characteristic frequency naturally arises for each of $x$ normal and $y$ normal surfaces: $\omega_{RC,x} = \left(\rho_{ct,x} C_{surf,x}\right)^{-1}$ and $\omega_{RC,y} = \left(\rho_{ct,y} C_{surf,y}\right)^{-1}$, where $C_{surf,x} = C_{surf}(\mathbf{e}_x)$ and $C_{surf,y} = C_{surf}(\mathbf{e}_y)$. While various scaling strategies could be employed depending on the characteristics to be studied, the local impedance functions are scaled by $\rho_{ct,x}$, and $Z_p$ is scaled by $\left(\rho_{ct,x}/8l_y\right)$ in this article. Notice that since the system is defined in 2D domain, the surface integral in Equation (17) has a length scale. The particle impedance is a harmonic average of the local impedance over $x$ and $y$ normal surfaces. Thus, the dimensionless particle impedance becomes

$$\tilde{Z}_p = \left(\frac{8l_y}{\rho_{ct,x}}\right)Z_p = \left(\frac{1}{2}\int_0^1 \left(\tilde{z}_{tot}|_{\tilde{x}=1}\right)^{-1} d\tilde{y} + \frac{1}{2}\gamma\int_0^1 \left(\tilde{z}_{tot}|_{\tilde{y}=1}\right)^{-1} d\tilde{x}\right)^{-1} \quad (23)$$

where $\tilde{z}_{tot} = z_{tot}/\rho_{ct,x}$ is the dimensionless local total impedance, and $\gamma = l_x/l_y$ is the geometric aspect ratio of the particle. Expanding $\tilde{z}_{tot}$ according to Equation (16) and performing the integrals, an analytical expression for $\tilde{Z}_p$ is obtained. Detailed algebraic and calculus steps are given in Appendix C.

$$\tilde{Z}_p^{-1} = \left(\frac{i\tilde{\omega}}{2}\right)\left(\frac{1}{\chi_x} + \frac{\gamma}{\nu\tau\chi_y}\right)$$
$$+ \frac{1}{2}\sum_{k=1}^{\infty}\Gamma_k B_k \left(\cos(\lambda_k) + \frac{\left(\frac{\gamma\Lambda_k}{\nu\lambda_k}\right)\sinh(\Lambda_k)\sin(\lambda_k) - \beta_y \sinh(\Lambda_k)\cos(\lambda_k)}{\Lambda_k \beta_y \cosh(\Lambda_k) + \Lambda_k^2 \sinh(\Lambda_k)}\right) \quad (24)$$

where $\nu = \rho_{cty}/\rho_{ctx}$, $\chi_x = \omega_{RC,x}/\omega_{D,x}$, $\chi_y = \omega_{RC,y}/\omega_{D,y}$, and $\gamma = l_x/l_y$ are additionally identified as dimensionless numbers governing the system response. Therefore, in total, eight dimensionless numbers are found to determine the behavior of the particle impedance. One of them, $\tilde{\omega}$, may be considered as an experimental variable as the applied frequency is gradually altered during an impedance measurement.

Table 1 shows typical values of the parameters, and the corresponding characteristic scales and dimensionless numbers, for Li ion batteries. Those values are adopted for the isotropic reference setup in examining $\tilde{Z}_p$ under various anisotropies. To investigate the effects of anisotropies in diffusion, insertion kinetics and surface accumulation, we may employ values different from the reference for some corresponding parameters. Since the reference values represent an isotropic setup, the diffusivities and the lengths in $x$ and $y$ directions are the same, respectively, and the charge transfer resistances and the surface capacitances on $x$ and $y$ normal surfaces are the same, respectively. Therefore, $\nu$, $\tau$ and $\gamma$ have values of unity. The charge transfer resistances are comparable to the diffusion characteristic resistances, so that $\beta_x$ and $\beta_y$ have values near unity. It implies that the RC semicircle and the Warburg regime have a similar size in the complex plan plot of $\tilde{Z}_p$ when the reference values are employed. On the other hand, the RC characteristic frequencies are about five orders higher than the diffusion characteristic frequencies. Thus, $\chi_x$ and $\chi_y$ have large values, and we can expect the RC semicircle and the diffusion impedance will be well separated in plots of $\tilde{Z}_p$.

*3.1. Effects of Anisotropic Diffusion*

The effect of anisotropic diffusion can be examined by using the $\tilde{Z}_p$ solution in Equation (24) and the parameters in Table 1, and varying one of the diffusivity values. Figure 5 shows behavior of $\tilde{Z}_p$ under various extents of anisotropy in diffusion. $D_{ch,y}$ is gradually changed from $2^{-8}$ times to $2^8$ times of $D_{ch,x}$, while $D_{ch,x}$ is fixed at $1.0 \times 10^{-9} \text{ cm}^2/\text{s}$. As described in the complex plane plot, Figure 5 (a), when the diffusion becomes anisotropic, we can obtain various slopes in the Warburg regime. In contrast, the slope has a given value around $45°$ when the diffusion is isotropic. The Warburg regime becomes more resistive with a lower slope when diffusion in one direction becomes slower. On the other hand, it becomes more capacitive with a higher slope when diffusion in one direction becomes faster.

In the magnitude and the phase angle plots, Figure 5 (b), transitions still take place near $\tilde{\omega} \approx 1$ and $\tilde{\omega} \approx 10^5$ with little change in its shape or magnitude.

When the charge transfer resistances and the diffusion characteristic resistances are comparable, like in the above case where $\beta_x \approx 1$ and $\beta_y \approx 1$, the ion flux near a particle surface is determined together by the corresponding charge transfer resistance and diffusion characteristic resistance. Since we are changing $D_{ch,y}$ in Figure 5, let us focus on the transport of ions solely in $y$ direction by looking at a $x$ normal cross-section of the system at an arbitrary location. When the diffusion becomes slower, there is local accumulation of ions near $y$ normal surfaces. On the other hand, when the diffusion becomes faster, it results in negligible gradient of ion concentration along $y$ direction. It means ions tend to be distributed homogeneously along the direction as soon as they enter the system through $y$ normal surfaces. When this phenomenon is combined with the diffusion pathways in $x$ direction, the local accumulation of ions near $y$ normal surfaces gives the resistive contribution, and the homogeneous distribution along $y$ direction gives the capacitive contribution to the Warburg regime of $\tilde{Z}_p$. On the other hand in $x$ direction, the ion diffusion maintains typical Fick's behavior as long as $\beta_x \approx 1$. Regardless of the diffusion anisotropy, it is the diffusion in $x$ direction which causes the transition in the diffusion impedance in Figure 5. Thus, the transition from the Warburg-like regime to the capacitive regime is located near $\tilde{\omega} \approx 1$, or equivalently near $\omega \approx \omega_{D,x}$, at which the diffusion penetration depth in $x$ direction reaches the symmetry plane at $x = 0$. Also, the transition keeps a magnitude scale of $\rho_{D,x}$, even though $\rho_{D,y}$ varies by orders. In general, when diffusion in one or more directions has the characteristic resistance similar to the charge transfer resistances on the surfaces, it determines the transition frequency and the magnitude scale of diffusion impedance. Anisotropy in diffusion allows various phase angle in the Warburg-like regime.

In particular, when $D_{ch,y} \gg D_{ch,x}$ in the above case with $\nu \approx 1$ and $\beta_x \approx 1$, the Gerischer limit can be defined. In more general, the Gerischer limit can be found when the system has strong anisotropy in diffusion but has relatively even insertion rate. Additionally, the largest diffusion characteristic resistance and the charge transfer resistances should be similar in order. Under such conditions, the ion flux through the surface normal to the faster diffusion direction could be analogized by homogenous reactions that simultaneously taking place with 1D diffusion along the slower direction. Therefore, in the Gerischer limit, the impedance response of 2D diffusion and surface insertion can be approximated by a derivative of the 1D Gerischer impedance. Mathematically, the approximation can be performed by averaging the system variables over the cross-section normal the slower diffusion direction. Then, the transport problem in the system is reduced to a typical 1D diffusion-reaction problem, and its response can be obtained in an analytical form:

$$\tilde{Z}_{p,G}^{-1} = \left(\frac{i\tilde{\omega}}{2}\right)\left(\frac{1}{\chi_x} + \frac{\gamma}{\nu\tau\chi_y}\right) + \frac{1}{2}\left[\left(1 + \frac{\tau\beta_y}{i\tilde{\omega}}\right)\left(1 + \frac{\beta_x \coth(\zeta)}{\zeta}\right)\right]^{-1} \quad (25)$$

where $\tilde{Z}_{p,G}$ is the Gerischer limit of particle impedance, and $\zeta = \sqrt{i\tilde{\omega} + \tau\beta_y}$. As implied by its name, a bounded Gerischer impedance function, $\beta_x \coth(\zeta)/\zeta$, can be found in place of the diffusion impedance. Detailed derivation of the solution is shown in Appendix D.

Such trends, however, are not general but depend on the parameter values and the corresponding dimensionless numbers. If $\beta_x \gg 1$ and $\beta_y \gg 1$ for instance, the insertion reaction stays near its equilibrium, and the ions are supplied at the surface as quickly as they diffuse away. Therefore, the current to the system is largely determined by its diffusion rate regardless of the insertion kinetics, and the charge transfer resistance appears negligible compared to that of the diffusion impedance. In this case, whenever $D_{ch,x} \gg D_{ch,y}$ or $D_{ch,x} \ll D_{ch,y}$, the ion diffusion in the system approaches 1D diffusion behavior in the faster direction. Accordingly, the Warburg regime becomes straighter with $45°$ slope, compared to the Warburg regime of isotropic 2D diffusion that has a positive curvature. The magnitude and the transition frequency of the diffusion impedance are then determined by the diffusion in the faster direction. On the other hand, when $\beta_x \ll 1$ and $\beta_y \ll 1$, it is the ion insertion rate at the surface that determines the response of the system, and the ions diffuse quickly and evenly throughout the system as soon as inserted. In this case, contribution of the ion diffusion appears as a purely capacitive behavior at frequencies lower than the RC characteristic frequencies. As a result, anisotropy in diffusion does not affect the impedance behavior.

*3.2. Effects of Anisotropic Surface Kinetics*

Varying one of the charge transfer resistances, we can also study the effect of anisotropic surface kinetics. Figure 6 shows behavior of $\tilde{Z}_p$ under various extents of anisotropy in surface kinetics, using the parameter values in Table 1. Among the parameters, $\rho_{ct,y}$ is gradually changed from $2^{-8}$ times to $2^8$ times of $\rho_{ct,x}$, while $\rho_{ct,x}$ is fixed at 44.06 $\Omega\text{cm}^2$. In the complex plane plot, Figure 6 (a), the diameter of the RC semicircle represents the total surface resistance. As $\rho_{ct,y}$ changes, the total surface resistance varies accordingly; when $\rho_{ct,y}$ decreases, the total surface resistance shrinks together. However, when $\rho_{ct,y}$ increases, the diameter converges to a certain value, rather than diverging. At this limit, even though $y$ normal surface is becoming blocked, the current can still pass through $x$ normal surface, and the total resistance converges to $\rho_{ct,x}$, scaled by $l_x/(l_x+l_y)$. Basically, the total surface resistance is a harmonic average of the charge transfer resistances over the system surface, and it tends to converge to the smaller value, scaled by the corresponding aspect ratio, when the two charge transfer resistances differ significantly. The overall RC characteristic frequency, around which the RC semicircle transits from the capacitive to the resistive behavior, is inversely proportional to the total surface resistance and the total surface capacitance. Thus when $\rho_{ct,y}$ decreases, the transition takes place at higher frequencies in Figure 6 (b), but when $\rho_{ct,y}$ increases, the transition frequency does not change significantly.

There is one more interesting aspect in this result: the anisotropy in surface kinetics also affects the diffusion impedance. Whenever $\rho_{ct,y}$ decreases or increases, if the two charge transfer resistances differ significantly, the diffusion impedance approaches its 1D limit. As shown in the complex plane plot, Figure 6 (a), the Warburg regime becomes straighter and less steep at the both extremes. It indicates that diffusion in the system becomes effectively one dimensional when the two surfaces have significantly different charge transfer resistances. It is the diffusion normal to the more active, or less resistive, surface that dominates the ion transport in the system. For example, when $\rho_{ct,y} \gg \rho_{ct,x}$, the majority of ions come from $x$ normal surface and diffuse predominantly in $x$ direction. The diffusion impedance thus approaches its 1D limit which represents the response of the diffusion in $x$ direction.

*3.3. Effects of Anisotropic Surface Capacitance*

Likewise, employing the solution of $\tilde{Z}_p$ and varying one of the surface capacitances in Table 1, we can study the effect of anisotropic surface capacitance on the battery impedance behavior. Figure 7 shows behavior of $\tilde{Z}_p$ under various extent of anisotropy in surface capacitance. $C_{surf,y}$ is gradually changed from $2^{-8}$ times to $2^8$ times of $C_{surf,x}$, while $C_{surf,x}$ is fixed at $1.0 \times 10^{-5}\ \text{F}/\text{cm}^2$. As shown in the phase angle and the magnitude plots, Figure 7 (b), when $C_{surf,y}$ increases, the overall RC characteristic frequency shifts to a lower value. Accordingly, in the complex plane plot, Figure 7 (a), the RC semicircle tends to be convoluted with the diffusion impedance. On the other hand, the RC transition frequency changes by little when $C_{surf,y}$ decreases. The total surface capacitance is an arithmetic average of the surface capacitances weighted by the corresponding surface lengths. Therefore, when the two surface capacitances are significantly different, the larger one determines the total surface capacitance as long as the surface lengths are similar. The total surface capacitance increases together with $C_{surf,y}$ when $C_{surf,y}$ increases, but it converges to the constant $C_{surf,x}$ when $C_{surf,y}$ decreases. Accordingly, the overall RC characteristic frequency shifts to a lower value when $C_{surf,y}$ increases, but it stays around the reference value when $C_{surf,y}$ decreases.

## 4. Overall Impedance of an Electrode with Anisotropic 2D Rectangular Particles

While the particle impedance has been studied up to this point, we now expand our view to the overall impedance of a battery electrode. The overall electrode impedance is an integrative response of all the particles in an electrode. Therefore, by investigating the overall electrode impedance, we can study the effects of distributions in anisotropic properties among the particles. In this section, consider a porous battery electrode that has hypothetic 2D rectangular particles or rod-shaped particles with rectangular cross-sections, as shown in Figure 8. In both cases, behavior of charge carriers in each particle can be described using the particle model we studied in the previous section, shown in Figure 4. Thus, $\tilde{Z}_p$ solution in Equation (24) can represent the corresponding impedance response of each particle. For such electrodes, two of the apparent distributions are the length distributions of the particles or the cross-sections. As discussed in Section 2, when there are distributions in properties among the particles, the overall impedance can be represented by a harmonic average of the particle impedance over the distributions. Considering the two lengths are the distributed parameters of our concern, we can define continuous random variables that represent their distributions, and the overall electrode impedance become:

$$\tilde{Z}_{ov} = \left( \frac{8\bar{L}_y N_p}{\rho_{ct,x}} \right) Z_{ov} = \left( \int_0^\infty \int_0^\infty \Pr\nolimits_{\tilde{L}_x, \tilde{L}_y} \left( \tilde{l}_x, \tilde{l}_y \right) \tilde{Z}_p^{-1} \left( \tilde{l}_x, \tilde{l}_y \right) \tilde{l}_y d\tilde{l}_x d\tilde{l}_y \right)^{-1} \quad (26)$$

where $\tilde{Z}_{ov}$ is the dimensionless overall electrode impedance, which is obtained by scaling $Z_{ov}$ by $\left( \rho_{ct,x} / 8\bar{L}_y N_p \right)$. $L_x$ and $L_y$ are the random variables representing the length distributions in $x$ and $y$ directions, respectively, and $\tilde{L}_x = L_x / \bar{L}_x$ and $\tilde{L}_y = L_y / \bar{L}_y$ are the corresponding dimensionless random variables, scaled by their respective means, $\bar{L}_x$ and $\bar{L}_y$. Then, $\tilde{l}_x$ and $\tilde{l}_y$ are the realizations of $\tilde{L}_x$ and $\tilde{L}_y$, respectively, and $\Pr\nolimits_{\tilde{L}_x, \tilde{L}_y}$ is the joint PDF of $\tilde{L}_x$ and $\tilde{L}_y$. $\tilde{Z}_p \left( \tilde{l}_x, \tilde{l}_y \right)$ is the dimensionless particle impedance we studied in the previous section, which is funtionalized in terms of $\tilde{l}_x$ and $\tilde{l}_y$. We employed a bivariate log-normal PDF to describe the distribution in $\tilde{L}_x$ and $\tilde{L}_y$. The distribution may have various spread, skewness, and correlation, and it would have different influence on the behavior of $\tilde{Z}_{ov}$. While we are focusing on the distributions in length scales here, the general approach in this article can be used to study distribution effects of any anisotropic parameters.

*4.1. Effect of uncorrelated length distributions*

From now on, we suppose that the particles or the cross-sections have strongly anisotropic diffusion and surface kinetics. They are considered to have faster diffusion in $x$ direction and faster insertion kinetics on $x$ normal surface. Therefore, we set $D_{ch,x} = 20 D_{ch,y}$ and $\rho_{ct,x} = 1/40\, \rho_{ct,y}$, while fixing $D_{ch,y}$ and $\rho_{ct,y}$ at their reference values: $D_{ch,y} = 1.0 \times 10^{-9}\, \text{cm}^2/\text{s}$ and $\rho_{ct,y} = 1.0 \times 10^{-5}\, \text{F/cm}^2$. For the other parameters, their reference values in Table 1 are employed as well. For a single particle or a cross-section, the agiler surface kinetics on $x$ normal surface and the faster diffusion in $x$ direction makes the majority of ions intercalate through $x$ normal surface and diffuse mainly in $x$ direction. In other words, the ion transport is strongly aligned in $x$ direction and can be approximated by 1D transport in the direction. Hence, $\tilde{l}_x$ becomes the main diffusion length and $\tilde{l}_y$ becomes the secondary diffusion length. In the particle impedance, then, the total surface resistance will converge to 2, or $\rho_{ct,x}/4l_y$ in dimensional quantity. Also, the particle diffusion impedance will have a characteristic resistance of $\beta_x$, or $\rho_{D,x}/\rho_{ct,x}$, and will transitions near $\tilde{\omega} \approx 1$, or $\omega \approx \omega_{D,x}$, at which the diffusion penetration depth in the main diffusion direction reaches the center of the particle or the cross-section

The random variables representing the length distributions, $\tilde{L}_x$ and $\tilde{L}_y$, may have different distributions. Figure 9 (a) shows scattered plots of their various distributions employed in calculating $\tilde{Z}_{ov}$, where $\Sigma_{xx}$, $\Sigma_{yy}$, $\Sigma_{xy} = \Sigma_{yx}$ are the variance of $\tilde{L}_x$, the variance of $\tilde{L}_y$ and the covariance of $\tilde{L}_x$ and $\tilde{L}_y$. Each distribution is a bivariate log-normal distribution, and $\tilde{L}_x$ and $\tilde{L}_y$ are considered independent from each other at this time $(\Sigma_{xy} = \Sigma_{yx} = 0)$. Figure 9 (b) shows the complex plane plot of $\tilde{Z}_{ov}$ examined with the length distributions in Figure 9 (a). Only the distribution in main diffusion length, $\tilde{L}_x$, affects the impedance behavior, while the distribution in secondary diffusion length, $\tilde{L}_y$, has little effect. As the distribution in $\tilde{L}_x$ becomes broader with increasing $\Sigma_{xx}$, the transition in diffusion impedance becomes smoother. Simultaneously, the capacitive regime starts deviating from the vertical behavior and shows a CPE-like behavior. Such trend is not affected by the distribution in $\tilde{L}_y$ or by the value of $\Sigma_{yy}$, as long as the length distributions are independent from each other. This is because ion intercalation and diffusion in each particle is nearly one-dimensional in $x$ direction. Therefore, distribution in $\tilde{L}_x$ leads to dispersion in the transition frequency of diffusion impedance and makes the transition spread over a wider frequency range. This is very similar to the effect of size distribution of isotropic particles on diffusion impedance, which is discussed extensively in Reference 17. On the other hand, the secondary diffusion length does not affect the ion transport behavior in the particle, and its distribution has negligible influence on $\tilde{Z}_{ov}$.

*4.2. Effect of correlated length distributions*

It becomes a different story when the length distributions are correlated. Now we need to consider the fact that, in the rectangular geometry of the particle or the cross-section, the secondary diffusion length, $\tilde{l}_y$, determines the effective surface area, or length, over which the majority of ions are being inserted. Thus, $\tilde{l}_y$ can also be defined as the main insertion length, and it gives weighting on the impedance response in the main diffusion direction. When $\tilde{l}_x$ and $\tilde{l}_y$ are positively correlated, long diffusion lengths are likely to be paired with long insertion lengths. Figure 10 (a) shows scattered plots of correlated distributions with various values of correlation, $\rho_{xy}$, and constant $\Sigma_{xx}$ and $\Sigma_{yy}$. As the distributions are correlated more with a larger value of $\rho_{xy}$, it becomes even more likely that large $\tilde{l}_x$ and large $\tilde{l}_y$ are paired together, and small $\tilde{l}_x$ and small $\tilde{l}_y$ are paired together. Therefore, response of a long diffusion length is likely to be weighted more by a long insertion length in the overall electrode impedance. Figure 10 (b) shows the behavior of $\tilde{Z}_{ov}$ in a complex plane, using the distributions in Figure 10 (a). As $\rho_{xy}$ increases, the capacitive regime in diffusion impedance becomes more resistive and shifts in the positive real direction. This is because the heavier weighting on the response of long diffusion lengths presents resistive contribution to the overall diffusion impedance when the length distributions are correlated.

    Unlike the behavior of $\tilde{Z}_{ov}$ under uncorrelated length distributions, it is affected by the distribution in secondary diffusion length, or main insertion length, when the length distributions are correlated. Figure 11 (a) shows scattered plots of correlated length distributions with various $\Sigma_{yy}$ and constant $\Sigma_{xx}$ and $\rho_{xy}$. Figure 11 (b) shows the behavior of $\tilde{Z}_{ov}$ on a complex plane, using the distributions in Figure 11 (a). As $\tilde{L}_y$ is distributed wider with a larger value of $\Sigma_{yy}$, the capacitive regime in diffusion impedance becomes more resistive and shifts in the positive real direction. In fact, the effect of increasing the correlation and the effect of increasing the variance in main insertion length appear very similar, because both basically give heavier weighting on the response of long diffusion lengths. Given a non-zero correlation, when $\tilde{L}_y$ has a wider distribution, large $\tilde{l}_x$ is likely

to be paired with even larger $\tilde{l}_y$ and small $\tilde{l}_x$ is likely to be paired with even smaller $\tilde{l}_y$, compared to their pairs when $\tilde{L}_y$ has a narrower distribution. Thus, the response of long diffusion lengths is weighted even more as $\Sigma_{yy}$ increases. It provides more resistive contribution to the diffusion impedance, and makes the capacitive regime shift in the positive real direction.

## 5. Conclusion

Although anisotropic materials play a crucial role in advanced batteries for energy storage, impedance models for a battery electrode with anisotropic particles are rarely studied. In this paper, we therefore developed an impedance model for a battery electrode that has anisotropic active particles. Using the model, the particle impedances were examined with various anisotropies in particle properties. It was found that the diffusion impedance may have various slopes in the Warburg regime when ion diffusion in the particle is anisotropic. The Warburg regime becomes more capacitive when one of the diffusivities increases, and it becomes more resistive when one of the diffusivities decreases. This trend results from the well-balanced resistance scales, and it is not observable when the charge transfer resistances and the diffusion characteristic resistances are significantly different from each others. On the other hand, when surface insertion kinetics is orientation-dependent, the total surface resistance of a particle tends to converge to the smaller charge transfer resistance of the more active surface, scaled by the corresponding surface area. The diffusion impedance is also affected and approaches its 1D limit, because ions mainly diffuse in the direction normal to the more active surface. Lastly, when the surfaces have different values of capacitance, the total surface capacitance of a particle tends to converge to the larger one, scaled by the corresponding surface area. Therefore, the RC element in particle impedance transitions at significantly lower frequencies when one of the surface capacitances increases, but the transition stays around the similar frequency range when one of the surface capacitance decreases.

Additionally, we expanded our view and examined the overall electrode impedance to study effects of particle length distributions. It is assumed that the active particles in the electrode have strong anisotropies both in surface kinetics and ion diffusion, which make ions effectively intercalate and diffuse in one direction. The main diffusion length and the main insertion length were defined, which determines the diffusion length and the insertion area, respectively, for the majority of ions. When the two lengths are distributed independently, only the distribution in main diffusion length affects the diffusion impedance, making it transition smoother and its capacitive regime deviate from the purely vertical behavior. If the length distributions are correlated, it is likely that the response of long diffusion lengths is weighted more by long insertion lengths. As a result, the distribution in main insertion length gives resistive contribution to the diffusion impedance. The resistive contribution becomes more significant when the correlation is higher or when the variance in main insertion length is larger. Using the general model presented in this article, it is also possible to investigate distribution effects of other anisotropic particle properties.

**Acknowledgements**

This work was partially supported by a grant from the Samsung-MIT Alliance and by a fellowship to JS from the Kwanjeong Educational Foundation.

**Appendix A. Nomenclature**

| | |
|---|---|
| $A_p$ | particle surface area |
| $c$ | local ion concentration |
| $c_s$ | surface ion concentration |
| $(\nabla c)_s$ | ion concentration gradient at surface |
| $\hat{\tilde{c}}$ | $= \left(-\partial \Delta \phi_{eq}/\partial c\right) \hat{c} / \Delta \hat{\phi}$, dimensionless local ion concentration |
| $C_{surf}$ | surface capacitance |
| $C_{surf,x}$ | $= C_{surf}(\mathbf{e}_x)$, surface capacitance on $x$ normal surface |
| $C_{surf,y}$ | $= C_{surf}(\mathbf{e}_y)$, surface capacitance on $y$ normal surface |
| $\mathbf{D}_{ch}$ | chemical diffusivity tensor |
| $D_{ch,x}$ | chemical diffusivity in $x$ direction |
| $D_{ch,y}$ | chemical diffusivity in $y$ direction |
| $D_{ch,z}$ | chemical diffusivity in $z$ direction |
| $e$ | elementary electric charge |
| $\mathbf{e}_x$ | unit vector in $x$ direction |
| $\mathbf{e}_y$ | unit vector in $y$ direction |
| $i$ | $=\sqrt{-1}$, unit imaginary number |
| $j_0$ | exchange current density |
| $j_{acc}$ | accumulation current density |
| $j_{ins}$ | insertion current density |
| $j_{tot}$ | total current density |
| $J_p$ | total current on particle surface |
| $J_{ov}$ | overall electrode current |
| $k$ | Boltzmann's constant |
| $l_x$ | one half of particle length in $x$ direction |
| $l_y$ | one half of particle length in $y$ direction |
| $\tilde{l}_x$ | $= l_x/\overline{L}_x$, dimensionless particle length in $x$ direction |
| $\tilde{l}_y$ | $= l_y/\overline{L}_y$, dimensionless particle length in $y$ direction |
| $L_x$ | one half of particle length in $x$ direction, a random variable |
| $L_y$ | one half of particle length in $y$ direction, a random variable |
| $\overline{L}_x$ | mean value of $L_x$ |
| $\overline{L}_y$ | mean value of $L_y$ |
| $\tilde{L}_x$ | $= L_x/\overline{L}_x$, dimensionless particle length in $x$ direction, a random variable |
| $\tilde{L}_y$ | $= L_y/\overline{L}_y$, dimensionless particle length in $y$ direction, a random variable |
| **n** | surface normal vector |
| $n_p$ | index number of each active particle |
| $N_p$ | total number of active particles |
| $\Pr_{\mathbf{V}_d}$ | joint probability density function of $\mathbf{V}_d$ |
| $\Pr_{\tilde{L}x,\tilde{L}y}$ | joint probability density function of $\tilde{L}_x$ and $\tilde{L}_y$ |
| $t$ | time variable |
| $T$ | temperature |
| $\mathbf{v}_d$ | vector of distributed parameters, a realization |
| $\mathbf{V}_d$ | vector of distributed parameters, a random vector variable |
| $x$ | spatial variable in $x$ direction |
| $\tilde{x}$ | $= x/l_x$, dimensionless spatial variable in $x$ direction |
| $X$ | arbitrary variable |
| $X_0$ | reference state response in $X$ |
| $X_1$ | $\varepsilon$-order perturbation in $X$ |
| $\hat{X}$ | Fourier coefficient of perturbation in $X$ |
| $y$ | spatial variable in $y$ direction |
| $\tilde{y}$ | $= y/l_y$, dimensionless spatial variable in $y$ direction |
| $z_{acc}$ | local accumulation impedance |
| $z_D$ | local diffusion impedance |
| $z_{ins}$ | local insertion impedance |
| $z_{tot}$ | local total impedance |
| $Z_p$ | particle impedance |
| $Z_{ov}$ | overall electrode impedance |
| $\tilde{z}_{tot}$ | $= z_{tot}/\rho_{ct,x}$, dimensionless local total impedance |
| $\tilde{Z}_p$ | $= 8l_y Z_p/\rho_{ct,x}$, dimensionless particle impedance |
| $\tilde{Z}_{p,G}$ | Gerischer limit of dimensionless particle impedance |
| $\tilde{Z}_{ov}$ | $= 8\bar{L}_y N_p Z_{ov}/\rho_{ct,x}$, dimensionless overall electrode impedance |

*Greek letters*

| | |
|---|---|
| $\alpha$ | charge transfer coefficient |
| $\beta_x$ | $= \rho_{D,x}/\rho_{ct,x}$, ratio of diffusion characteristic resistance in $x$ direction and charge transfer resistance on $x$ normal surface |
| $\beta_y$ | $= \rho_{D,y}/\rho_{ct,y}$, ratio of diffusion characteristic frequency in $y$ direction and charge transfer resistance on $y$ normal surface |
| $\chi_x$ | $= \omega_{RC,x}/\omega_{D,x}$, ratio of RC characteristic frequency on $x$ normal surface and diffusion characteristic frequency in $x$ direction |
| $\chi_y$ | $= \omega_{RC,y}/\omega_{D,y}$, ratio of RC characteristic frequency on $y$ normal surface and diffusion characteristic frequency in $y$ direction |
| $\varepsilon$ | arbitrary small number |
| $\Delta\phi$ | potential drop across electrolyte/active material interface |
| $\Delta\phi_{eq}$ | equilibrium potential drop of insertion reaction |
| $-\partial\Delta\phi_{eq}/\partial c$ | Nernst shift |
| $\gamma$ | $= l_x/l_y$, geometric aspect ratio of a rectangular particle |
| $\eta$ | surface overpotential |
| $\nu$ | $= \rho_{cty}/\rho_{ctx}$, ratio of charge transfer resistances |
| $\rho_{xy}$ | correlation between $\tilde{L}_x$ and $\tilde{L}_y$ |
| $\rho_{ct}$ | $= kT/j_0 e$, charge transfer resistance |

| | | |
|---|---|---|
| $\rho_{ct,x}$ | $= \rho_{ct}(\mathbf{e}_x)$, charge transfer resistance on $x$ normal surface | |
| $\rho_{ct,y}$ | $= \rho_{ct}(\mathbf{e}_y)$, charge transfer resistance on $y$ normal surface | |
| $\rho_{D,x}$ | $= \left(-\partial \Delta\phi_{eq}/\partial c\right) l_x / eD_{ch,x}$, diffusion characteristic resistance in $x$ direction | |
| $\rho_{D,y}$ | $= \left(-\partial \Delta\phi_{eq}/\partial c\right) l_y / eD_{ch,y}$, diffusion characteristic resistance in $y$ direction | |
| $\Sigma_{xx}$ | variance in $\tilde{L}_x$ | |
| $\Sigma_{yy}$ | variance in $\tilde{L}_y$ | |
| $\Sigma_{xy}$, $\Sigma_{yx}$ | covariance of $\tilde{L}_x$ and $\tilde{L}_y$ | |
| $\tau$ | $= \omega_{D,y}/\omega_{D,x}$, ratio of diffusion characteristic frequencies | |
| $\omega$ | applied frequency | |
| $\omega_{D,x}$ | $= D_{ch,x}/l_x^2$, diffusion characteristic frequency in $x$ direction | |
| $\omega_{D,y}$ | $= D_{ch,y}/l_y^2$, diffusion characteristic frequency in $y$ direction | |
| $\omega_{RC,x}$ | $= \left(\rho_{ct,x} C_{surf,x}\right)^{-1}$, RC characteristic frequency on $x$ normal surface | |
| $\omega_{RC,y}$ | $= \left(\rho_{ct,y} C_{surf,y}\right)^{-1}$, RC characteristic frequency on $y$ normal surface | |
| $\tilde{\omega}$ | $= \omega/\omega_{D,x}$, dimensionless applied frequency | |
| $\Omega$ | domain of $\mathbf{V}_d$ | |

**Appendix B. Solution to the 2D anisotropic transport problem[64]**

Ion diffusion in an anisotropic 2D rectangular domain is governed by Equations (21). This boundary value PDE problem can be reformulated with homogeneous boundary conditions, by shifting the solution by a unity.

$$\Theta = \hat{\tilde{c}} + 1 \qquad (B.1)$$

Having homogeneous boundary conditions not only makes subsidiary calculations simpler, but also ensures that termwise differentiation of its trigonometric Fourier series solution converges to the real derivative. This is not always true if nonhomogeneous boundary conditions were included.

As depicted in Figure B.1, the governing PDE and boundary conditions for $\Theta$ become

$$
\begin{aligned}
i\tilde{\omega}(\Theta - 1) &= \frac{\partial^2 \Theta}{\partial \tilde{x}^2} + \tau \frac{\partial^2 \Theta}{\partial \tilde{y}^2} \\
\left.\frac{\partial \Theta}{\partial \tilde{x}}\right|_{\tilde{x}=1} &+ \beta_x \Theta\big|_{\tilde{x}=1} = 0 \\
\left.\frac{\partial \Theta}{\partial \tilde{y}}\right|_{\tilde{y}=1} &+ \beta_y \Theta\big|_{\tilde{y}=1} = 0 \\
\left.\frac{\partial \Theta}{\partial \tilde{x}}\right|_{\tilde{x}=0} &= 0 \\
\left.\frac{\partial \Theta}{\partial \tilde{y}}\right|_{\tilde{y}=0} &= 0
\end{aligned}
\qquad (B.2)
$$

Notice that the boundary conditions in Equations (B.2) are all homogeneous. We approach this problem using the FFT method, where an exact solution is found by expanding the solution in terms of a set of known basis functions and determining the unknown coefficient functions in the expansion. To begin with, we hypothesize that the solution $\Theta$ can be written in a series expansion as

$$\Theta(\tilde{x}, \tilde{y}) = \sum_{k}^{\infty} \Phi_k(\tilde{y}) \Psi_k(\tilde{x}) \tag{B.3}$$

where $\Psi_k$ are the basis functions, and $\Phi_k$ are the coefficient functions. $\Psi_k$ are the solutions to a certain eigenvalue problem, and also called its eigenfunctions. The eigenvalue problem of concern here is

$$\begin{aligned} \frac{d^2\Psi}{d\tilde{x}^2} &= -\lambda^2 \Psi \\ \left.\frac{d\Psi}{d\tilde{x}}\right|_{\tilde{x}=1} + \beta_x \Psi|_{\tilde{x}=1} &= 0 \\ \left.\frac{d\Psi}{d\tilde{x}}\right|_{\tilde{x}=0} &= 0 \end{aligned} \tag{B.4}$$

where $\lambda$ may have only certain values, termed eigenvalues. The general solution for the eigenvalue problem is

$$\Psi = A \sin(\lambda \tilde{x}) + B \cos(\lambda \tilde{x}) \tag{B.5}$$

where $A$ and $B$ are the constant coefficients. Applying boundary conditions, $A$ and $B$ must satisfy the following equation:

$$\begin{pmatrix} \lambda \cos(\lambda) + \beta_x \sin(\lambda) & \beta_x \cos(\lambda) - \lambda \sin(\lambda) \\ \lambda & 0 \end{pmatrix} \begin{pmatrix} A \\ B \end{pmatrix} = \begin{pmatrix} 0 \\ 0 \end{pmatrix} \tag{B.6}$$

While it turns out $A = 0$, the algebraic equation has its nontrivial solution only when the determinant of the coefficient matrix is zero. This requirement leads to the characteristic equation of the eigenvalue problem which determines possible values of $\lambda$.

$$\lambda(\beta_x \cos(\lambda) - \lambda \sin(\lambda)) = 0 \tag{B.7}$$

If $\lambda = 0$, the corresponding eigenfunction becomes a trivial function, $\Psi = 0$, and the zero eigenvalue is excluded. Therefore for $k = 1, 2, 3, \ldots$, the eigenvalue $\lambda_k$ is given implicitly by

$$\lambda_k \tan(\lambda_k) = \beta_x \tag{B.8}$$

where $0 < \lambda_k < \lambda_{k+1}$, and the corresponding eigenfunction $\Psi_k$ is given by

$$\Psi_k(\tilde{x}) = B_k \cos(\lambda_k \tilde{x}) \tag{B.9}$$

This set of eigenfunctions is an orthogonal set according to the Sturm-Liouville theory, and the coefficient $B_k$ can be determined to make the set even orthonormal:

$$\langle \Psi_k, \Psi_k \rangle = \int_0^1 \left( B_k \cos(\lambda_k \tilde{x}) \right)^2 d\tilde{x} = 1 \tag{B.10}$$

for $k = 1, 2, 3, \ldots$. Notice that the weighting function of the eigenvalue problem is a unity. Therefore, $B_k$ is given by

$$B_k = 2\sqrt{\frac{\lambda_k}{2\lambda_k + \sin(2\lambda_k)}} \tag{B.11}$$

In the FFT method, the boundary value PDE problem for $\Theta$ is transformed to boundary value ordinary differential equation (ODE) problems for $\Phi_k$ for $k = 1, 2, 3, \ldots$, which is defined by Equation (B.3) or equivalently by

$$\Phi_k(\tilde{y}) = \langle \Psi_k(\tilde{x}), \Theta(\tilde{x}, \tilde{y}) \rangle = \int_0^1 \Psi_k(\tilde{x}) \Theta(\tilde{x}, \tilde{y}) d\tilde{x} \tag{B.12}$$

The ODE and boundary conditions governing each $\Phi_k$ are obtained by transforming the original set of equations, Equations (B.2), in this problem. They are transformed by taking every terms in the equations inner product with $\Psi_k$. For instance, the governing equation can be transformed for $\Phi_k$ as:

$$i\tilde{\omega}\langle\Psi_k,(\Theta-1)\rangle = \left\langle\Psi_k,\frac{\partial^2\Theta}{\partial\tilde{x}^2}\right\rangle + \tau\left\langle\Psi_k,\frac{\partial^2\Theta}{\partial\tilde{y}^2}\right\rangle \qquad (B.13)$$

It can be reduced to a second order ODE for $\Phi_k$.

$$\frac{d^2\Phi_k}{d\tilde{y}^2} - \left(\frac{i\tilde{\omega}+\lambda_k^2}{\tau}\right)\Phi_k + \frac{i\tilde{\omega}}{\tau}\left(\frac{B_k\sin(\lambda_k)}{\lambda_k}\right) = 0 \qquad (B.14)$$

Likewise, the boundary conditions at $\tilde{y}=1$ and $\tilde{y}=0$ are also transformed to new boundary conditions for $\Phi_k$.

$$\begin{aligned}\left.\frac{d\Phi_k}{d\tilde{y}}\right|_{\tilde{y}=1} + \beta_y\,\Phi_k\big|_{\tilde{y}=1} &= 0 \\ \left.\frac{d\Phi_k}{d\tilde{y}}\right|_{\tilde{y}=0} &= 0\end{aligned} \qquad (B.15)$$

The general solution for this ODE problem is

$$\Phi_k(\tilde{y}) = P_k\sinh(\Lambda_k\tilde{y}) + Q_k\cosh(\Lambda_k\tilde{y}) + \Gamma_k \qquad (B.16)$$

where $P_k$ and $Q_k$ are the constant coefficients, and

$$\Lambda_k = \sqrt{\frac{i\tilde{\omega}+\lambda_k^2}{\tau}} \qquad (B.17)$$

$\Gamma_k$ is a specific solution of the ODE problem, and we take a constant form of

$$\Gamma_k = \left(\frac{i\tilde{\omega}}{i\tilde{\omega}+\lambda_k^2}\right)\left(\frac{B_k\sin(\lambda_k)}{\lambda_k}\right) \qquad (B.18)$$

Applying the boundary conditions in Equations (B.15),

$$\Phi_k(\tilde{y}) = \Gamma_k \left(1 - \frac{\beta_y \cosh(\Lambda_k \tilde{y})}{\beta_y \cosh(\Lambda_k) + \Lambda_k \sinh(\Lambda_k)}\right) \tag{B.19}$$

The overall solution for $\Theta(\tilde{x}, \tilde{y})$ then can be obtained by plugging the final expressions for $\Psi_k(\tilde{x})$ and $\Phi_k(\tilde{y})$ into its Fourier series expansion.

$$\Theta(\tilde{x}, \tilde{y}) = \sum_{k=1}^{\infty} \Gamma_k \left(1 - \frac{\beta_y \cosh(\Lambda_k \tilde{y})}{\beta_y \cosh(\Lambda_k) + \Lambda_k \sinh(\Lambda_k)}\right) B_k \cos(\lambda_k \tilde{x}) \tag{B.20}$$

where $\lambda_k$, $B_k$, $\Lambda_k$ and $\Gamma_k$ are defined above. Finally we can shift $\Theta$ back by a negative unity to have the solution for $\hat{\tilde{c}}$.

$$\hat{\tilde{c}}(\tilde{x}, \tilde{y}) = -1 + \sum_{k=1}^{\infty} \Gamma_k \left(1 - \frac{\beta_y \cosh(\Lambda_k \tilde{y})}{\beta_y \cosh(\Lambda_k) + \Lambda_k \sinh(\Lambda_k)}\right) B_k \cos(\lambda_k \tilde{x}) \tag{B.21}$$

This solution is employed in the main article and Appendix C, to calculate various impedance functions.

### Appendix C. Particle impedance of the 2D rectangular particle

Particle impedance, $\tilde{Z}_p$, is a harmonic average of total local impedance, $\tilde{z}_{tot}$, over the particle surface. For the 2D rectangular particle, as expanded in Equation (23) in the main article, the subsidiary calculus can be performed separately on $x$ normal and $y$ normal surfaces. Looking at the integral $\tilde{z}_{tot}$ inverse of on $x$ normal surface first,

$$\int_0^1 \left(\tilde{z}_{tot}\big|_{\tilde{x}=1}\right)^{-1} d\tilde{y} = \int_0^1 \left(\left(1 + \tilde{z}_D\big|_{\tilde{x}=1}\right)^{-1} + \left(\tilde{z}_{acc}\big|_{\tilde{x}=1}\right)^{-1}\right) d\tilde{y} \tag{C.1}$$

where $\tilde{z}_{tot} = z_{tot}/\rho_{ct,x}$, and the other dimensionless local impedance functions in the equation can be written as following:

$$\tilde{z}_D\big|_{\tilde{x}=1} = \frac{z_D\big|_{\tilde{x}=1}}{\rho_{ct,x}} = \left(\frac{\partial \Delta\phi_{eq}}{\partial c}\right)\frac{\hat{c}\big|_{\tilde{x}=1}}{\rho_{ct,x}\hat{j}_{ins}\big|_{\tilde{x}=1}} = \frac{-\hat{\tilde{c}}\big|_{\tilde{x}=1}}{\hat{\tilde{c}}\big|_{\tilde{x}=1}+1} \qquad (C.2)$$

and

$$\tilde{z}_{acc}\big|_{\tilde{x}=1} = \frac{z_{acc}\big|_{\tilde{x}=1}}{\rho_{ct,x}} = \frac{1}{i\omega C_{surf,x}\rho_{ct,x}} = \frac{\chi_x}{i\tilde{\omega}} \qquad (C.3)$$

where $\chi_x = \omega_{RC,x}/\omega_{D,x}$. Plugging Equations (C.2) and (C.3) in Equation (C.1), we can analytically perform the integral.

$$\int_0^1 \left(\tilde{z}_{tot}\big|_{\tilde{x}=1}\right)^{-1} d\tilde{y} = \int_0^1 \left(\Theta\big|_{\tilde{x}=1} + i\tilde{\omega}/\chi_x\right) d\tilde{y}$$
$$= \left(\frac{i\tilde{\omega}}{\chi_x}\right) + \sum_{k=1}^{\infty} \Gamma_k \left(1 - \frac{\beta_y \sinh(\Lambda_k)}{\Lambda_k \beta_y \cosh(\Lambda_k) + \Lambda_k^2 \sinh(\Lambda_k)}\right) B_k \cos(\lambda_k) \qquad (C.4)$$

where $\lambda_k$, $B_k$, $\Lambda_k$, and $\Gamma_k$ are defined in Appendix B for $k = 1, 2, 3, \dots$.

The integral of $\tilde{z}_{tot}$ inverse on $y$ normal surface could be obtained through the similar method. Compared to those on $x$ normal surface, components of $\tilde{z}_{tot}$ on $y$ normal surface contain a prefactor of $\nu = \left(\rho_{ct,y}/\rho_{ct,x}\right)$ when they are scaled by $\rho_{ct,x}$.

$$\int_0^1 \left(\tilde{z}_{tot}\big|_{\tilde{y}=1}\right)^{-1} d\tilde{x} = \int_0^1 \left(\left(\nu + \tilde{z}_D\big|_{\tilde{y}=1}\right)^{-1} + \left(\tilde{z}_{acc}\big|_{\tilde{y}=1}\right)^{-1}\right) d\tilde{x} \qquad (C.5)$$

where

$$\tilde{z}_D\big|_{\tilde{y}=1} = \frac{z_D\big|_{\tilde{y}=1}}{\rho_{ct,x}} = \left(\frac{\rho_{ct,y}}{\rho_{ct,x}}\right)\left(\frac{\partial \Delta\phi_{eq}}{\partial c}\right)\frac{\hat{c}\big|_{\tilde{y}=1}}{\rho_{ct,y}\hat{j}_{ins}\big|_{\tilde{y}=1}} = \nu\left(\frac{-\hat{\tilde{c}}\big|_{\tilde{y}=1}}{\hat{\tilde{c}}\big|_{\tilde{y}=1}+1}\right) \quad (C.6)$$

and

$$\tilde{z}_{acc}\big|_{\tilde{y}=1} = \frac{z_{acc}\big|_{\tilde{y}=1}}{\rho_{ct,x}} = \frac{1}{i\omega C_{surf,y}\rho_{ct,x}} = \nu\tau\frac{\chi_y}{i\tilde{\omega}} \quad (C.7)$$

where $\tau = \omega_{D,y}/\omega_{D,x}$ and $\chi_y = \omega_{RC,y}/\omega_{D,y}$. As a result, the integral of $\tilde{z}_{tot}$ inverse on $y$ normal surface becomes

$$\int_0^1 \left(\tilde{z}_{tot}\big|_{\tilde{y}=1}\right)^{-1} d\tilde{x} = \nu^{-1}\int_0^1 \left(\Theta\big|_{\tilde{y}=1} + (i\tilde{\omega}/\tau\chi_y)\right) d\tilde{x}$$
$$= \frac{1}{\nu}\left(\left(\frac{i\tilde{\omega}}{\tau\chi_y}\right) + \sum_{k=1}^{\infty} \Gamma_k \left(\frac{\Lambda_k \sinh(\Lambda_k)}{\beta_y \cosh(\Lambda_k) + \Lambda_k \sinh(\Lambda_k)}\right)\left(\frac{B_k \sin(\lambda_k)}{\lambda_k}\right)\right) \quad (C.8)$$

Putting the results together into Equation (23), the particle impedance has the following form:

$$\tilde{Z}_p^{-1} = \left(\frac{i\tilde{\omega}}{2}\right)\left(\frac{1}{\chi_x} + \frac{\gamma}{\nu\tau\chi_y}\right)$$
$$+ \frac{1}{2}\sum_{k=1}^{\infty} \Gamma_k B_k \left(\cos(\lambda_k) + \frac{\left(\frac{\gamma\Lambda_k}{\nu\lambda_k}\right)\sinh(\Lambda_k)\sin(\lambda_k) - \beta_y \sinh(\Lambda_k)\cos(\lambda_k)}{\Lambda_k \beta_y \cosh(\Lambda_k) + \Lambda_k^2 \sinh(\Lambda_k)}\right) \quad (C.9)$$

This solution is employed in the main article to evaluate $\tilde{Z}_p$ and other related impedance functions.

**Appendix D. The Gerischer limit of particle impedance**

The Gerischer limit can be defined when the system has strongly anisotropic diffusion but relatively even surface insertion rate. It is also required that the charge transfer resistances are comparable to the largest diffusion characteristic resistance. In our system, such conditions are satisfied when $D_{ch,y} \gg D_{ch,x}$, $v \approx 1$, and $\beta_x \approx 1$. Under these conditions, the ion concentration varies much less in $y$ direction compared to that in $x$ direction. It indicates that we can approximate the concentration as a function of $x$ only. Given that the concentration field is approximately one-dimensional in $x$ direction, the local value can be replaced by the cross-sectional average along the $y$ direction, which is defined as:

$$\hat{\bar{\tilde{c}}}(\tilde{x}) = \int_0^1 \hat{\tilde{c}}(\tilde{x}, \tilde{y}) d\tilde{y} \tag{D.1}$$

where $\hat{\bar{\tilde{c}}}$ is the cross-sectional average dimensionless concentration at a certain $x$. To obtain the governing equation for $\hat{\bar{\tilde{c}}}$, we average each term of the anisotropic diffusion equation, in Equations (21), over the cross section normal to $x$ direction.

$$\frac{d^2 \hat{\bar{\tilde{c}}}}{d\tilde{x}^2} = \left(i\tilde{\omega} + \tau\beta_y\right)\hat{\bar{\tilde{c}}} + \tau\beta_y \tag{D.2}$$

where we have approximated the surface concentration at $\tilde{y} = 1$ by the cross-sectional average concentration. As we reduce the dimension of the governing equation by cross-sectional averaging, we obtain two terms in place of the second order derivative. Each of them functions like a first order reaction and a zeroth order reaction in the governing equation. We can therefore expect that the impedance response under this condition would contain a Gerischer impedance element, may be modified by additional terms due to the zeroth order reaction term.

Appropriate boundary conditions can be found by averaging the boundary conditions at $\tilde{x} = 1$ and $\tilde{x} = 0$ in a similar manner.

$$-\beta_x^{-1} \left.\frac{d\hat{\tilde{c}}}{d\tilde{x}}\right|_{\tilde{x}=1} = 1 + \left.\hat{\tilde{c}}\right|_{\tilde{x}=1}$$
$$\left.\frac{d\hat{\tilde{c}}}{d\tilde{x}}\right|_{\tilde{x}=0} = 0$$
(D.3)

The general solution that satisfies Equation (D.2) is

$$\hat{\tilde{c}} = C\sinh(\zeta\tilde{x}) + D\cosh(\zeta\tilde{x}) - \left(\frac{\tau\beta_y}{i\tilde{\omega}+\tau\beta_y}\right) \tag{D.4}$$

where $C$ and $D$ are constants still unknown, and

$$\zeta = \sqrt{i\tilde{\omega}+\tau\beta_y} \tag{D.5}$$

The unknown constants can be determined by applying the boundary conditions. According to the symmetry condition in Equations (D.3), $C$ turns out to be zero. The other unknown constant, $D$, is then determined by the Robin boundary condition in Equations (D.3):

$$D = -\left(\frac{\beta_x}{\zeta\sinh(\zeta)+\beta_x\cosh(\zeta)}\right)\left(\frac{i\tilde{\omega}}{i\tilde{\omega}+\tau\beta_y}\right) \tag{D.6}$$

Adopting the definitions in the main article, the particle impedance in the Gerischer limit becomes

$$\tilde{Z}_{p,G} = \left(\frac{1}{2}\int_0^1 \left(\tilde{\tilde{z}}_{tot}\big|_{\tilde{x}=1}\right)^{-1} d\tilde{y} + \frac{1}{2}\gamma\int_0^1 \left(\tilde{\tilde{z}}_{tot}\big|_{\tilde{y}=1}\right)^{-1} d\tilde{x}\right)^{-1}$$
$$= \left(\frac{1}{2}\left(\tilde{\tilde{z}}_{tot}\big|_{\tilde{x}=1}\right)^{-1} + \frac{1}{2}\gamma\left(\tilde{\tilde{z}}_{acc}\big|_{\tilde{y}=1}\right)^{-1}\right)^{-1}$$
(D.7)

where $\tilde{Z}_{p,G}$ and $\tilde{\tilde{z}}_{tot}$ are the dimensionless particle impedance and the dimensionless total local impedance in the Gerischer limit, respectively; they are defined as like $\tilde{Z}_p$ and $\tilde{z}_{tot}$ with the same scaling strategy, except using the cross-sectional average variables in place of the corresponding original variables. Likewise, the other average impedance functions in the Gerischer limit are defined as like their correspondences, using cross-sectional average variables. Notice in $\tilde{Z}_{p,G}$ formula above that the insertion impedance on $y$ normal surface is not gone away, but transplanted in the insertion impedance on $x$ normal surface. $\tilde{\bar{z}}_{tot}$ is not a function of $y$, and the integrals are straightforward.

Using the definition of $z_{tot}$ in Equation (16),

$$\tilde{\bar{z}}_{tot} = \left( \left(1+\tilde{\bar{z}}_D\right)^{-1} + \left(\tilde{z}_{acc}\right)^{-1} \right)^{-1} \tag{D.8}$$

where the diffusion impedance can be calculated using the solution in Equation (D.4).

$$\tilde{\bar{z}}_D = \beta_x \frac{\left. \hat{\bar{c}} \right|_{\tilde{x}=1}}{\left. \left( d\hat{\bar{c}} / d\tilde{x} \right) \right|_{\tilde{x}=1}} = \frac{\beta_x}{\zeta} \coth(\zeta) + \left( \frac{\tau \beta_y}{i\tilde{\omega}} \right) \left( 1 + \frac{\beta_x \coth(\zeta)}{\zeta} \right) \tag{D.9}$$

Then, plugging Equations (D.8) and (D.9) into Equation (D.7), the cross-sectional average particle impedance becomes

$$\tilde{Z}_{p,G}^{-1} = \left( \frac{i\tilde{\omega}}{2} \right) \left( \frac{1}{\chi_x} + \frac{\gamma}{\nu \tau \chi_y} \right) + \frac{1}{2} \left[ \left( 1 + \frac{\tau \beta_y}{i\tilde{\omega}} \right) \left( 1 + \frac{\beta_x \coth(\zeta)}{\zeta} \right) \right]^{-1} \tag{D.10}$$

where $\beta_x \coth(\zeta)/\zeta$ is the bounded Gerischer impedance element. The additional term in the prefactor, $\tau \beta_y / i\tilde{\omega}$, is attributed to the last term in the governing equation, Equation (D.2), that functions like a zeroth order reaction term.

**Tables**

Table 1. Isotropic reference values of the parameters, characteristic scales and dimensionless numbers[20]

| Property | Value | Unit |
|---|---|---|
| $\left(-\partial \Delta \phi_{eq}/\partial c\right)$ | 20.27 | $Vcm^3/mol$ |
| $D_{ch,x}$, $D_{ch,y}$ | $1.0\times 10^{-9}$ | $cm^2/s$ |
| $C_{surf,x}$, $C_{surf,y}$ | $1.0\times 10^{-5}$ | $F/cm^2$ |
| $\rho_{ct,x}$, $\rho_{ct,y}$ | 44.06 | $\Omega cm^2$ |
| $l_x$, $l_y$ | $2.0\times 10^{-4}$ | cm |
| $\rho_{D,x}$, $\rho_{D,y}$ | 46.16 | $\Omega cm^2$ |
| $\omega_{D,x}$, $\omega_{D,y}$ | $2.5\times 10^{-2}$ | $s^{-1}$ |
| $\omega_{RC,x}$, $\omega_{RC,y}$ | $2.3\times 10^{3}$ | $s^{-1}$ |
| $\beta_x$, $\beta_y$ | 1.05 | -- |
| $\nu$ | 1 | -- |
| $\chi_x$, $\chi_y$ | $9.08\times 10^{4}$ | -- |
| $\tau$ | 1 | -- |
| $\gamma$ | 1 | -- |

**Figures**

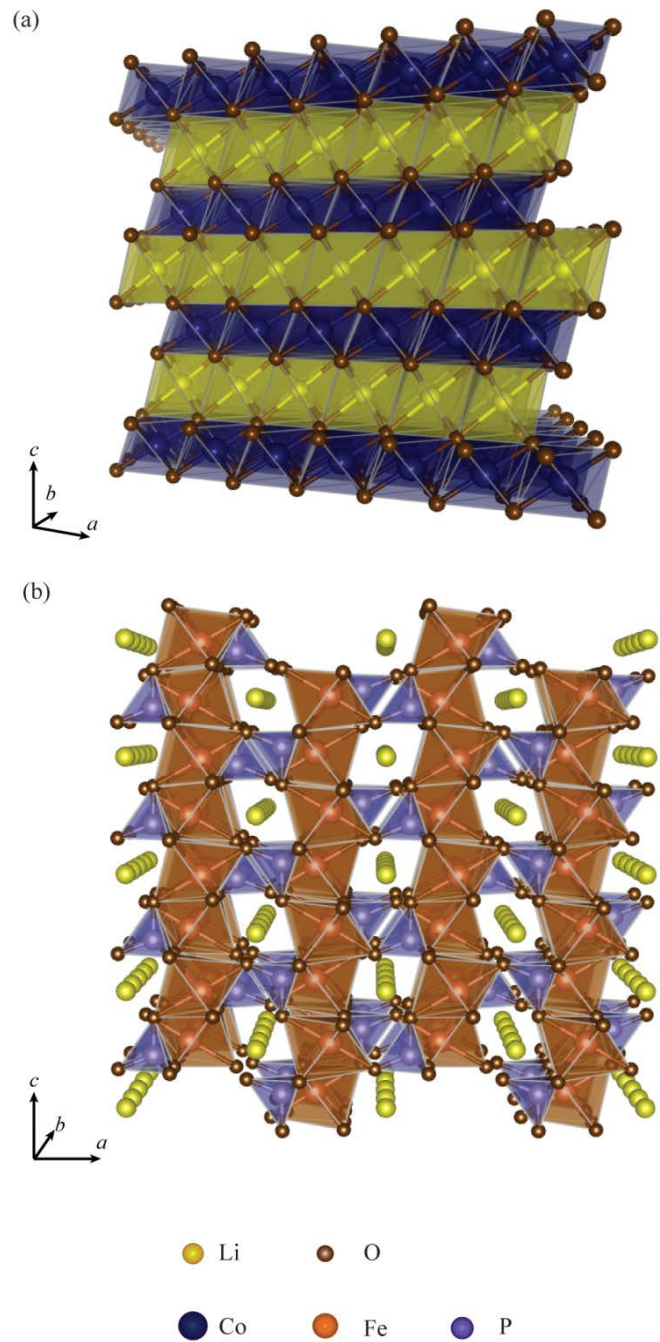

**Fig. 1.** Crystal structures of Li ion battery materials: (a) LiCoO2, a layered transition metal oxide structure, and (b) LiFePO4, an olivine structure[45]

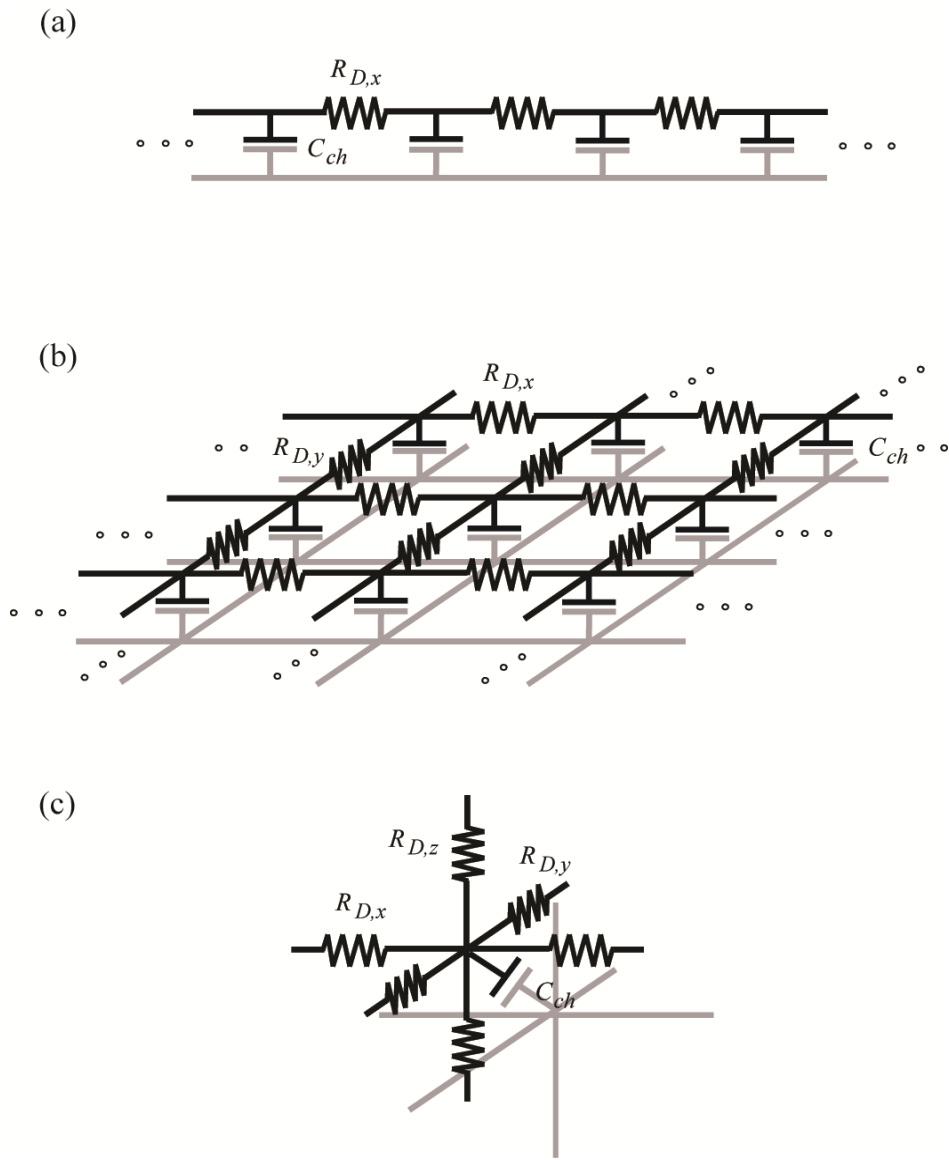

**Fig. 2.** Transmission line models describing transport of charge carriers in active material: (a) a 1D transmission line model, (b) a 2D transmission line model, and (c) a repeating node of a 3D transmission line model

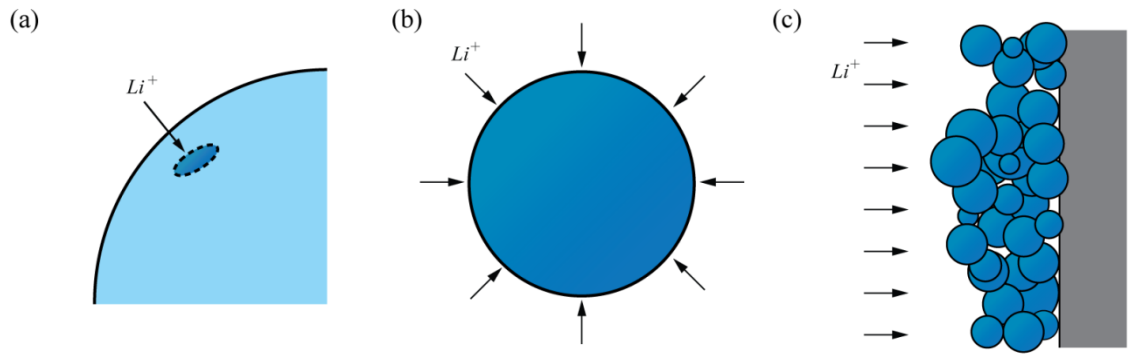

**Fig. 3.** Schematic drawing of the corresponding domains for (a) local impedance functions, (b) particle impedance, and (c) overall electrode impedance

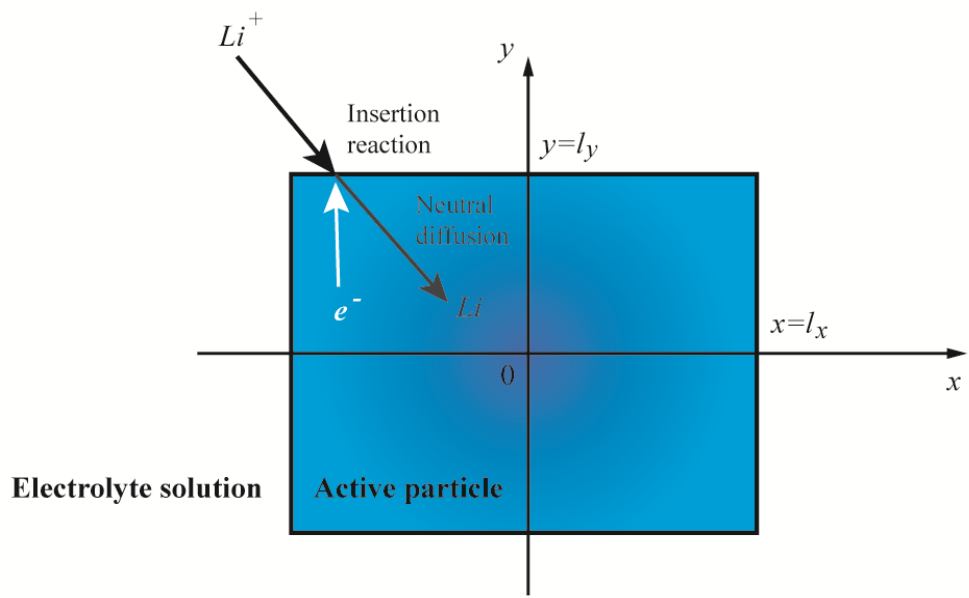

**Fig. 4.** Anisotropic 2D rectangular particle in electrolyte solution, and behavior of charge carriers in the system

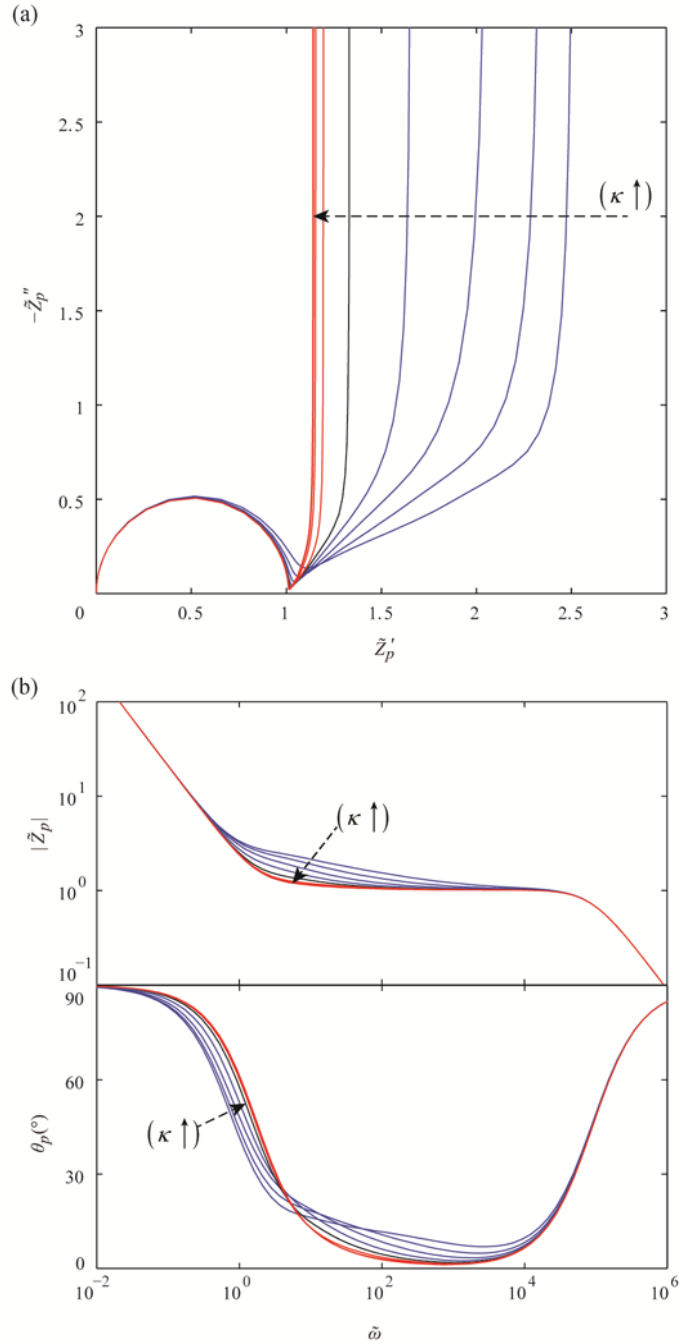

**Fig. 5.** Particle impedance varying the extent of anisotropy in diffusion: (a) complex plane plot, (b) magnitude and phase angle plots, which are also called Nyquist plot and Bode plots, respectively. In the figure, $D_{ch,y} = 2^{\kappa} D_{ch,x}$ for $\kappa = -8, -6, -4, -2, 0, 2, 4, 6, 8$, while $D_{ch,x}$ was fixed at $1.0 \times 10^{-9}\ \mathrm{cm^2/s}$. Curves are colored blue when $\kappa < 0$, black when $\kappa = 0$, and red when $\kappa > 0$.

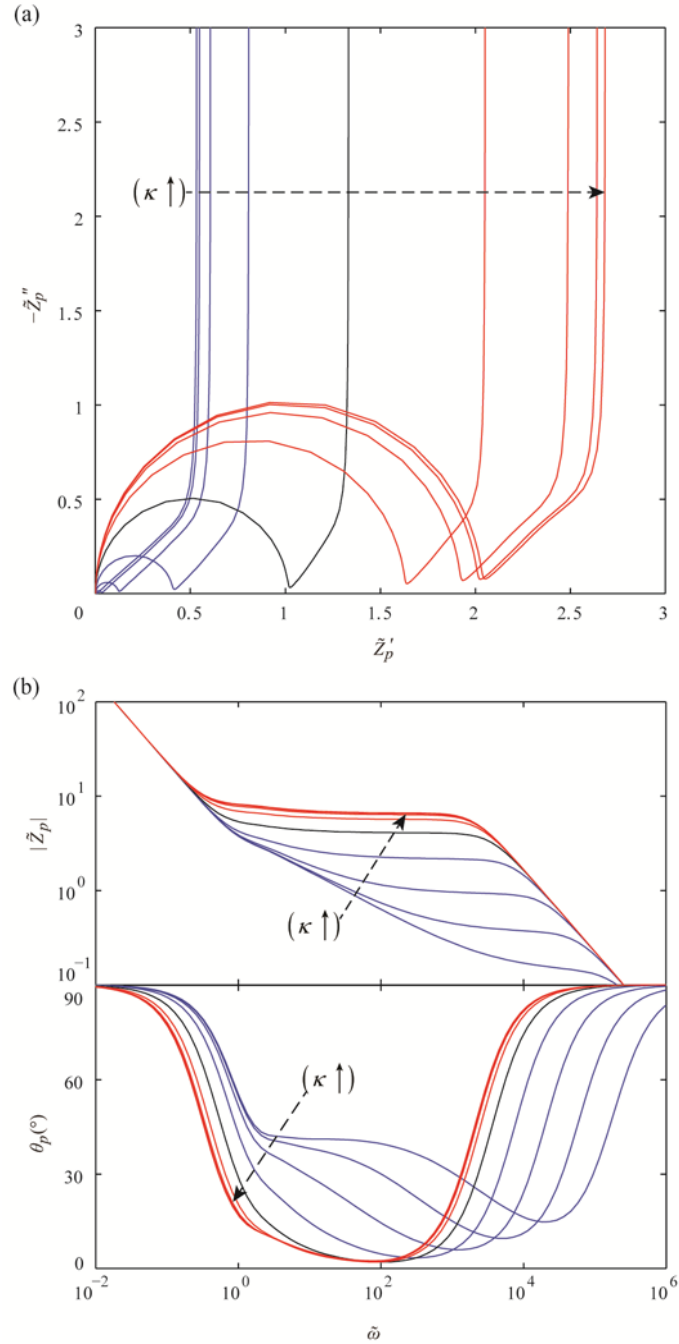

**Fig. 6.** Particle impedance varying the extent of anisotropy in surface kinetics: (a) complex plane plot, (b) magnitude and phase angle plots. In the figure, $\rho_{ct,y} = 2^{\kappa}\rho_{ct,x}$ for $\kappa = -8, -6, -4, -2, 0, 2, 4, 6, 8$, while $\rho_{ct,x}$ was fixed at $44.06 \Omega\text{cm}^2$. Curves are colored blue when $\kappa < 0$, black when $\kappa = 0$, and red when $\kappa > 0$.

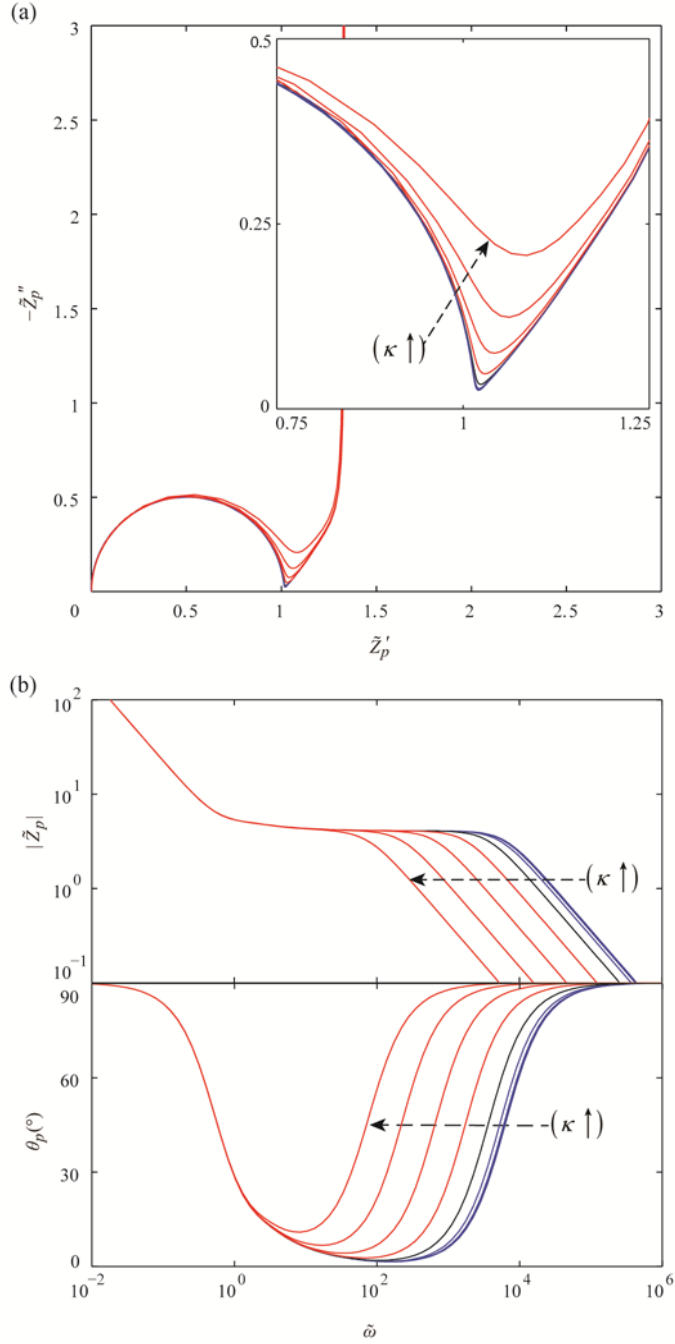

**Fig. 7.** Particle impedance varying the extent of anisotropy in surface capacitance: (a) complex plane plot, (b) magnitude and phase angle plots. In the figure, $C_{surf,y} = 2^{\kappa} C_{surf,x}$ for $\kappa = -8, -6, -4, -2, 0, 2, 4, 6, 8$, while $C_{surf,x}$ was fixed at $1.0 \times 10^{-5} \, \text{F/cm}^2$. Curves are colored blue when $\kappa < 0$, black when $\kappa = 0$, and red when $\kappa > 0$.

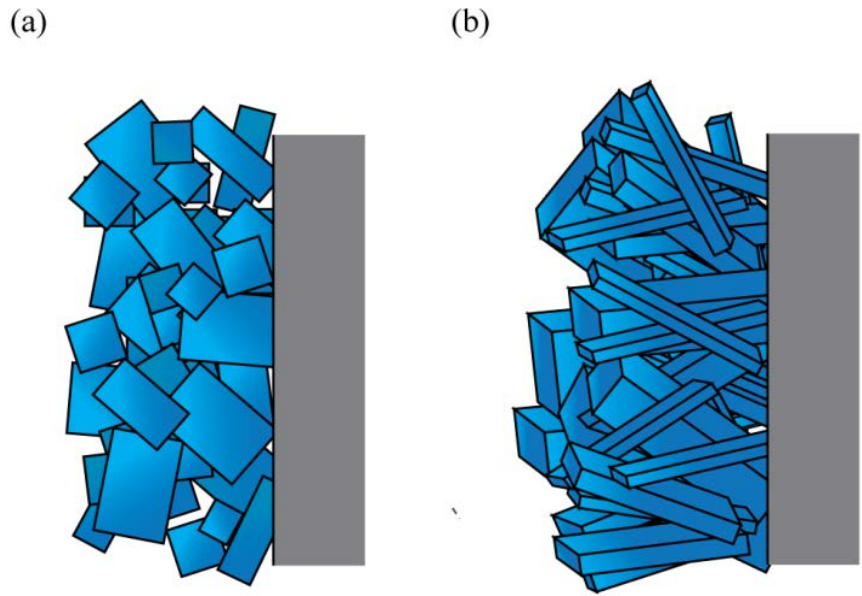

**Fig. 8.** Electrode configurations considered in Section 4: (a) an electrode with hypothetic 2D rectangular particles, and (b) an electrode with rod-like particles of rectangular cross-sections

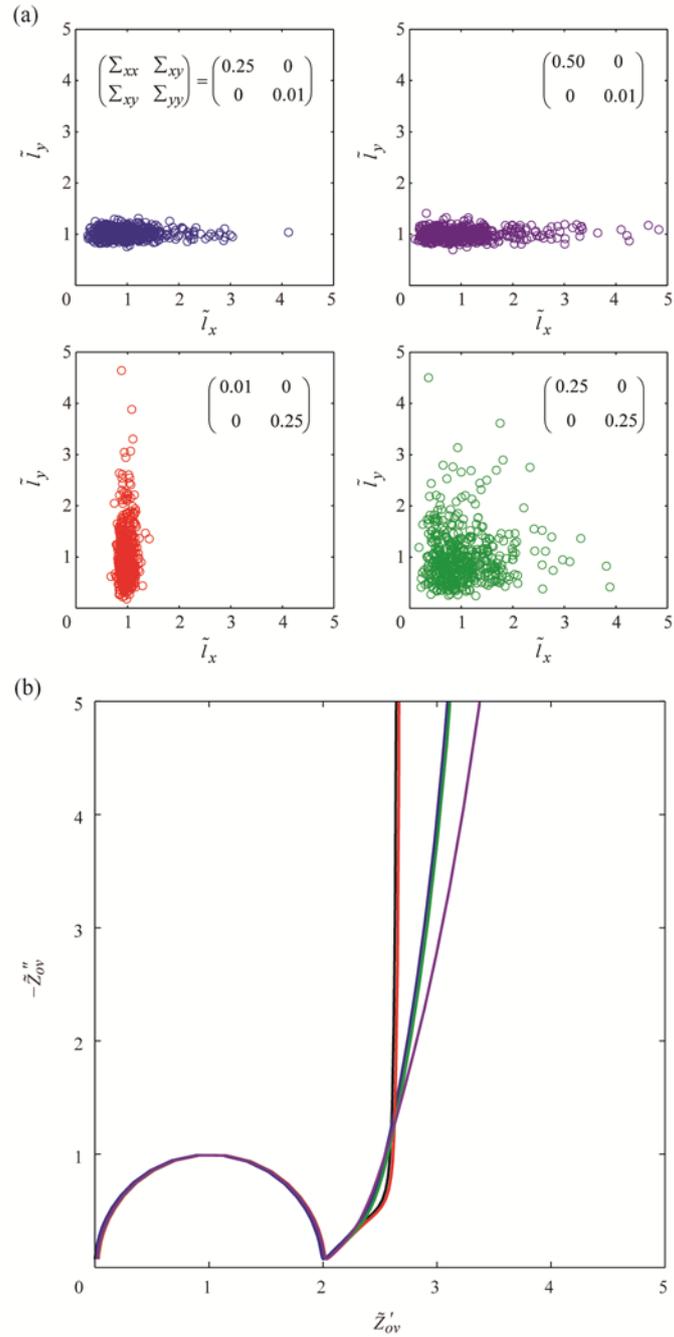

**Fig. 9.** (a) Uncorrelated length distributions with various covariance matrices, and (b) a complex plane plot of overall electrode impedance with the length distributions in (a). The distributions and the corresponding curves are paired by the same color, while the black curve represents the electrode impedance with identical particle lengths.

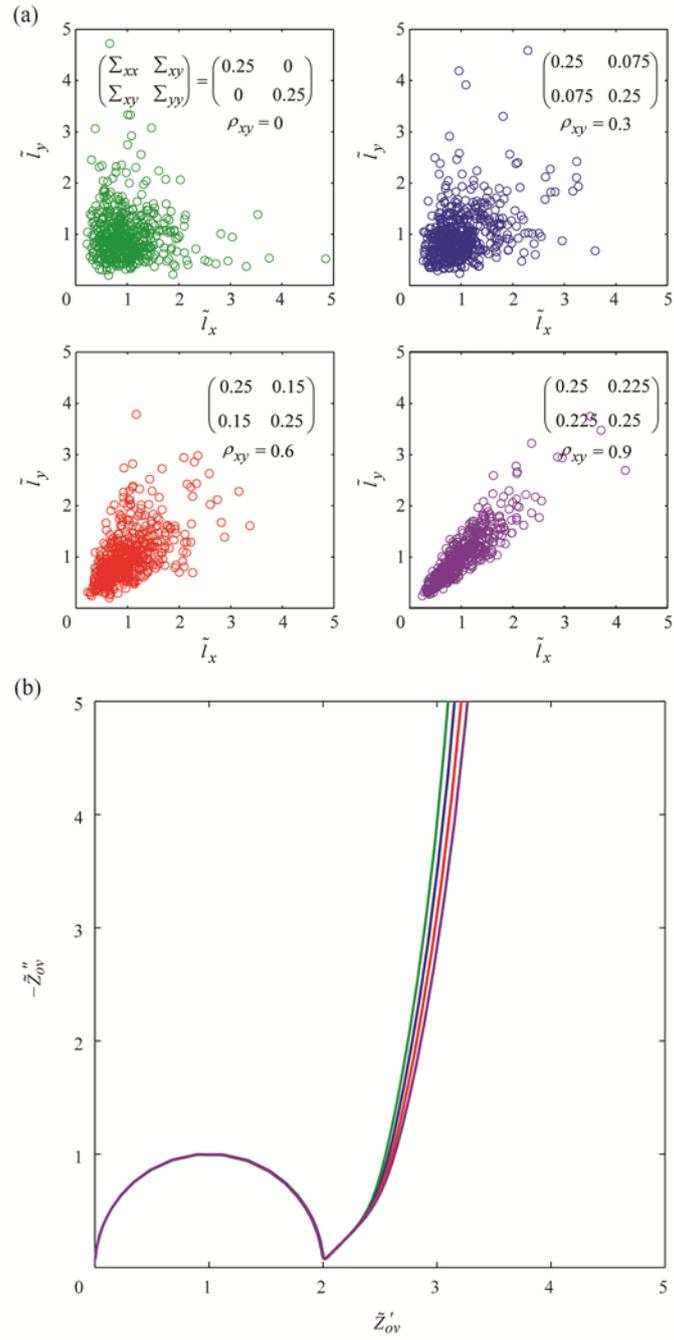

**Fig. 10.** (a) Correlated length distributions with various $\rho_{xy}$ values, and (b) a complex plane plot of overall electrode impedance with the length distributions in (a). The distributions and the corresponding curves are paired by the same color.

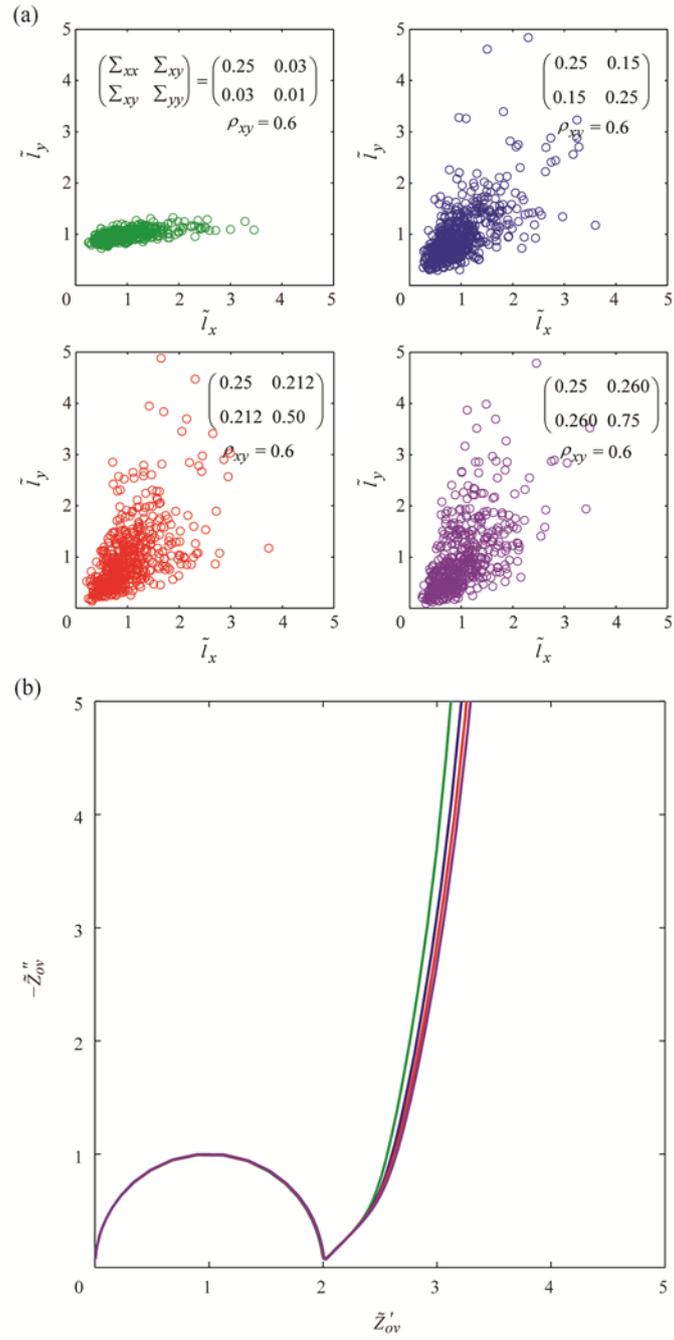

**Fig. 11.** (a) Correlated length distributions with various $\Sigma_{yy}$ values, and (b) a complex plane plot of overall electrode impedance with the length distributions in (a). The distributions and the corresponding curves are paired by the same color.

**Appendix Figures**

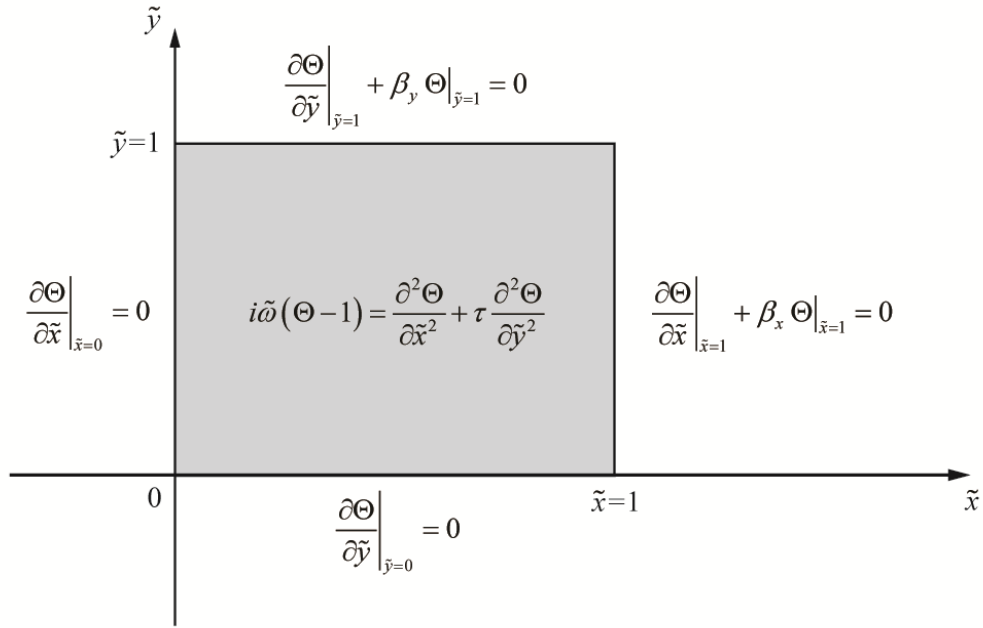

**Fig. B.1.** Dimensionless governing equation and boundary conditions for the 2D anisotropic ion transport problem